\documentclass[aps,pra,epsf,onecolumn,nofootinbib,superscriptaddress,showpacs,preprintnumbers,amsmath,amssymb,floatfix]{revtex4-1}
\usepackage{microtype}
\DisableLigatures[f]{encoding = *, family = *} 
\usepackage{amsmath,epsfig}
\usepackage{amssymb,amsfonts}
\usepackage{xcolor}
\usepackage{verbatim}
\usepackage{graphicx}
\usepackage{bm}
\usepackage{braket}
\usepackage{fouridx}
\usepackage[titletoc]{appendix}

\begin{document}
\title{Interplay of phase separation and itinerant magnetism for\\ correlated few fermions in a double-well}

\author{G.M. Koutentakis}
\affiliation{Center for Quantum Optical Technologies, University of Hamburg,
Department of Physics, Luruper Chaussee 149, 22761 Hamburg, Germany}
\affiliation{The Hamburg Centre for Ultrafast Imaging,
Universit\"{a}t Hamburg, Luruper Chaussee 149, 22761 Hamburg,
Germany}
\author{S.I. Mistakidis}
\affiliation{Center for Quantum Optical Technologies, University of Hamburg,
Department of Physics, Luruper Chaussee 149, 22761 Hamburg, Germany}
\author{P. Schmelcher}
\affiliation{Center for Quantum Optical Technologies, University of Hamburg,
Department of Physics, Luruper Chaussee 149, 22761 Hamburg, Germany}
\affiliation{The Hamburg Centre for Ultrafast Imaging,
Universit\"{a}t Hamburg, Luruper Chaussee 149, 22761 Hamburg,
Germany}

\date{\today}

\begin{abstract}
    We explore the stability of the phase separation phenomenon in few-fermion
    spin-$1/2$ systems confined in a double-well potential. It is shown that
    within the SU(2) symmetric case, where the total spin is conserved, the
    phase separation cannot be fully stabilized. An interaction regime characterized
    by metastable phase separation emerges for intermediate interactions which
    is inherently related with ferromagnetic spin-spin correlations
    emanating within each of the wells. The breaking of the SU(2) symmetry
    crucially affects the stability properties of the system as the
    phase separated state can be stabilized even for weak magnetic potential gradients.
    Our results imply an intricate relation between the phenomena of
    phase separation and ferromagnetism that lies beyond the
    view of the Stoner instability.
\end{abstract}
\maketitle

\section{Introduction}
         
Understanding the properties of itinerant magnetism has been a long-standing problem in condensed
matter physics \cite{Vollhardt,Brando}. Its importance stretches beyond this field of study since it
impacts the behaviour of a large class of quantum systems encountered e.g. in atomic physics \cite{GiorginiRev,Silverstein}. The
emergence of ferromagnetism in systems of spatially delocalized short-range repulsively interacting spinor fermions 
has been historically qualitatively understood in the framework of the
Stoner instability \cite{Stoner}. Within this framework ferromagnetism is related to the
phase separation of the different spin components and the formation of ferromagnetic domains
\cite{Grochowski,Salasnich,Salasnich1}. Ultracold atoms provide a fertile platform to investigate
such quantum many-body (MB) phenomena due to their exceptional tunability \cite{GiorginiRev}. Indeed, several experiments utilizing ensembles of
ultracold fermions have attempted to implement and study the Stoner instability
\cite{Ketterle1,Ketterle2,Valtolina,Scazza,LENS-Ketterle} but their results have been somewhat
inconclusive \cite{Pekker,Cui1}. 

The phase separation of Fermi systems has been studied in the case of strong attractive interactions
\cite{PhasSep1,PhasSep2,PhasSep3} where the phenomenon of spin-segregation for weak attraction
or repulsion has been identified \cite{Thomas1,Thomas2}. However, only recently experiments
attempted to address the relation between ferromagnetism and phase separation in the case of
a repulsively interacting Fermi-gas \cite{Valtolina,LENS-Ketterle}. For instance, it has been demonstrated \cite{Valtolina} that an
artificially prepared phase separated state becomes metastable for strong repulsions
which in turn implies the presence of a ferromagnetic instability. Accordingly, by
employing pump-probe spectroscopy the emergence of short-range two-body anti-correlations
in the repulsive Fermi-gas supporting some sort of ferromagnetic order has been revealed \cite{LENS-Ketterle},
while the possibility of macroscopic phase separation has been ruled out. These experimental  evidences
indicate that the relation between phase separation and magnetism might be more intricate and involved than it appears
within the framework of the Stoner instability manifested within the Hartree-Fock theory. Nevertheless, competing processes such as the Feshbach molecule
formation \cite{Chin} and its possible enhancement by coherent processes \cite{Pekker} have
hindered the experimental progress in this direction. As a consequence a complete understanding on how and 
via which mechanism phase separation and ferromagnetism are related remains still elusive.

Here we propose that one-dimensional (1D) few-body systems offer an ideal platform to provide
insight into these fundamental questions. Besides the suppression of the above-mentioned competing
processes which render the magnetic properties of 1D spin-$1/2$ fermions experimentally addressable
\cite{Zurn1,Zurn2}, the corresponding theoretical understanding of these properties is also 
advanced. Indeed, the availability of numerically-exact methods \cite{Lewenstein,Giann1,MLX} and the
development of powerful spin-chain models
\cite{Deuretzbacher,Zinnerchain,Levinsen,Yangchain,Cui3,Giann2,KMWR,Koutent} allows for the accurate
modeling of the magnetic properties emerging in 1D systems in the cases of strong
\cite{Deuretzbacher,Zinnerchain,Levinsen,Yangchain,Cui3,Giann2} and weak \cite{KMWR,Koutent}
interactions. Regarding the occurrence of phase separation previous studies revealed the role of the
breaking of the SU(2) symmetry, associated with the conservation of the total spin of the system.
Moreover, manifestations of the interplay between the magnetic properties and the phase separation
have also been reported \cite{Jenny,Jens,Cui2,Artem,KMWR,Koutent}. Below, we provide some
characteristic examples. It has been demonstrated \cite{Jenny,Jens} that contrary to mean-field
treatments phase separation does not occur during the interaction-quench dynamics of an SU(2)
symmetric system. However, the ground state of a system with weakly broken SU(2) symmetry is known
\cite{Cui2,Artem} to be phase separated in the case of infinite repulsion.  In contrast, it has been
shown that a parabolically confined initially spin-polarized Fermi-gas in the case of weak
interactions prefers a state of largely miscible spin components even when perturbed by a spin
dependent potential which weakly breaks the SU(2) symmetry \cite{KMWR}.
In particular, for sufficiently weak spin-dependent potentials a ferromagnetic order despite the
miscible character of the Fermi-gas has been established \cite{Koutent}.  However, a systematic
study that clarifies the relation between the phase separation and the magnetic properties of 1D
fermions unifying, also, the above results is currently absent. Furthermore, the comparison of the
underlying mechanisms provided by such a unification with the expectations of the Stoner instability
might provide invaluable insights into the study of magnetic phenomena emanating in more complex
systems.

Here we attempt to bridge this apparent gap in the literature by studying the stability of the phase
separated state during the correlated dynamics of fermionic ensembles confined in a double-well
(DW). The employed DW confinement allows for the experimental implementation of the phase separated
initial state \cite{Valtolina}.  This initial state is allowed to evolve for different values of the
interaction strength and the degree of the dynamical phase separation between the spin components is
monitored. To capture the correlated out-of-equilibrium dynamics of this spinor fermion system we
resort to the multilayer multiconfiguration time-dependent Hartree method for atomic mixtures
(ML-MCTDHX) \cite{MLX}. Focussing on an SU(2) invariant system and following the above-mentioned
procedure we find that for weak interactions the phase separation is unstable. While for increasing
repulsion an interaction regime where the phase separated state becomes metastable is unveiled.  To
identify the emergence of this metastable state and its relation with the magnetic properties of the
system we invoke an effective tight-binding model. The metastability of the phase separated state is
shown to be inherently connected with the appearance of a quasi-degenerate manifold of eigenstates
characterized by intra-well ferromagnetic correlations of both wells but a varying total spin. The
occurrence of this manifold is attributed to the ferromagnetic Hund exchange interactions
\cite{Hund1,Hund2,Hund3} emanating within each well of the DW setup. Moreover, the low-frequency tunneling
dynamics that leads to the decay of the metastable initial state provides a manifestation of the
antiferromagnetic Anderson kinetic exchange interactions \cite{Andersonkin}. These interactions 
act between the wells and result in the lifting of the degeneracy among states exhibiting intra-well
ferromagnetic correlations.

For larger interactions, the interband coupling introduced by
cradle-like processes \cite{Mistakidis1,Mistakidis2,Mistakidis3} is shown to result in a fastly
decaying dynamics of the phase separation, thus limiting the interaction regime where this
metastability of the initial state is exhibited.  The breaking of the SU(2) symmetry is found to
substantially affect the dynamics of the system. Indeed, the initial phase separated state of the
system can be stabilized by applying a linear magnetic potential gradient to the system. This
stabilization is much more prevalent in the case of intermediate interactions due to the occurrence
of quasi-degenerate eigenstates with different total spin. Our results demonstrate the relation of
the phase separation to the stability of the intra-well ferromagnetic order. Indeed, the interplay
of the Anderson and Hund exchange interactions is found to dictate the behaviour of the system in
terms of these two above phenomena implying that their relation is more intricate than what is
qualitatively expected in view of the Stoner instability.

This paper is structured as follows. In section II we introduce our setup and discuss its inherent
spin symmetries. Section III presents the MB dynamics of our system and showcases the important
features of the related eigenspectrum.  An effective tight-binding model of our system is introduced
in section IV which is subsequently utilized to expose the magnetic properties of the system during
the dynamics. In section V we study the dynamics in the case of a broken SU(2) symmetry. Finally, in
section VI we conclude and provide future perspectives.  In Appendices A and B we generalize our
results for more particles and different barrier heights respectively. Appendix C provides the
derivation of the Anderson effective kinetic exchange interaction for our DW setup and Appendix D
describes the employed numerical approach, namely the ML-MCTDHX method.

\section{Description of the system and relevant observables} 
\subsection{Hamiltonian and Symmetries}

We consider an interacting system consisting of $N$ spin-$1/2$ fermions of mass $m$ being confined
in an 1D DW trap. The latter is composed by a harmonic oscillator with frequency $\omega$ and a
Gaussian barrier.  Such a system is described by the MB Hamiltonian $\hat{H}=\hat H_\text{0}+\hat
H_{I}$, where $\hat{H}_0$ and $\hat H_I$ correspond to its non-interacting and interacting parts
respectively. The Hamiltonian, $\hat H$, expressed in harmonic oscillator units
($\hbar=m=\omega=1$), reads
\begin{equation}
    \hat H=\underbrace{\sum_{\alpha} \int {\rm d}x~ \hat{\psi}^\dagger_\alpha(x) \left(-\frac{1}{2}
    \frac{{\rm d}^2}{{\rm d}x^2} +\frac{1}{2} x^2 +V_0~e^{-\frac{x^2}{2
w^2}}\right)\hat{\psi}_{\alpha}(x)}_{\equiv \hat{H}_0}+ 
\underbrace{g \int {\rm d}x~\hat{\psi}^\dagger_\downarrow(x) \hat{\psi}^\dagger_\uparrow(x)
\hat{\psi}_\downarrow(x) \hat{\psi}_\uparrow(x)}_{\equiv \hat H_I},
\label{hamilt}
\end{equation}
where $\hat \psi_\alpha(x)$ denotes the fermionic field operator with spin-$\alpha \in
\{\uparrow,\downarrow\}$. $V_0$ and $w$ refer to the height and width of the Gaussian barrier
respectively. In the ultracold regime, $g$ describes the effective 1D $s$-wave contact interaction
strength between anti-aligned spins. This effective interaction strength, $g$, is known to be related
with the transverse confinement length and the 3D $s$-wave scattering length \cite{Olshanii}. The
above imply that the interaction strength is experimentally tunable with the aid of
confinement-induced and Fano-Feshbach resonances \cite{Chin}. The Hamiltonian of Eq. (\ref{hamilt})
is invariant under rotations in spin-space as it commutes with the total $\hat S_z$, $\hat S_\pm=\hat S_x
\pm i \hat S_y$ spin operators. The corresponding individual spin operators, $\hat S_k$, are defined as
\begin{equation}
    \hat S_k=\frac{1}{2}\int {\rm d}x~\sum_{\alpha,\beta}\hat{\psi}^\dagger_\alpha(x)\sigma^k_{\alpha,\beta}\hat{\psi}_\beta(x),
    \label{spin}
\end{equation}
with $\sigma^k$, $k \in \{x,y,z \}$, referring to the corresponding Pauli matrix. The system
additionally possesses an SU(2) symmetry since $\hat{H}$ [Eq. (\ref{hamilt})] commutes with the
total spin operator, $\hat S^2=\hat{S}_+\hat{S}_-+\hat{S}_z(\hat{S}_z-1)$. As we shall demonstrate
later on, this symmetry has a deep impact on the eigenspectrum of the system.

The behaviour of the single-particle Hamiltonian $\hat H_0$ for varying $V_0$ and $w$ is well-known
\cite{book,Thesis} and depicted in Fig. \ref{fig:setup} (a). For $V_0=0$ the harmonic oscillator
potential is retrieved and the single-particle spectrum consists of equidistant states. As $V_0$ is
increased, gradually all the eigenenergies, starting with the energetically two lowest ones, form
quasi-degenerate pairs of different parity states (herewith called bands). Employing linear
combinations of the two eigenstates forming the band, $b$, it is possible to construct the so-called
Wannier states, $\phi_{s}^b(x)$, which are localized either in the left, $s=L$ or the right well,
$s=R$ \cite{Wannier1}. The single-particle dynamics of a system initialized in such a Wannier state
is rather simple as the particle tunnels from each well to the other during the evolution with a
frequency given by the energy difference, $2 t^b$, between the two quasi-degenerate states  which
form the corresponding band.

\subsection{Initial State Characterization}
\begin{figure}[h]
    \centering
    \includegraphics[width=1.0\textwidth]{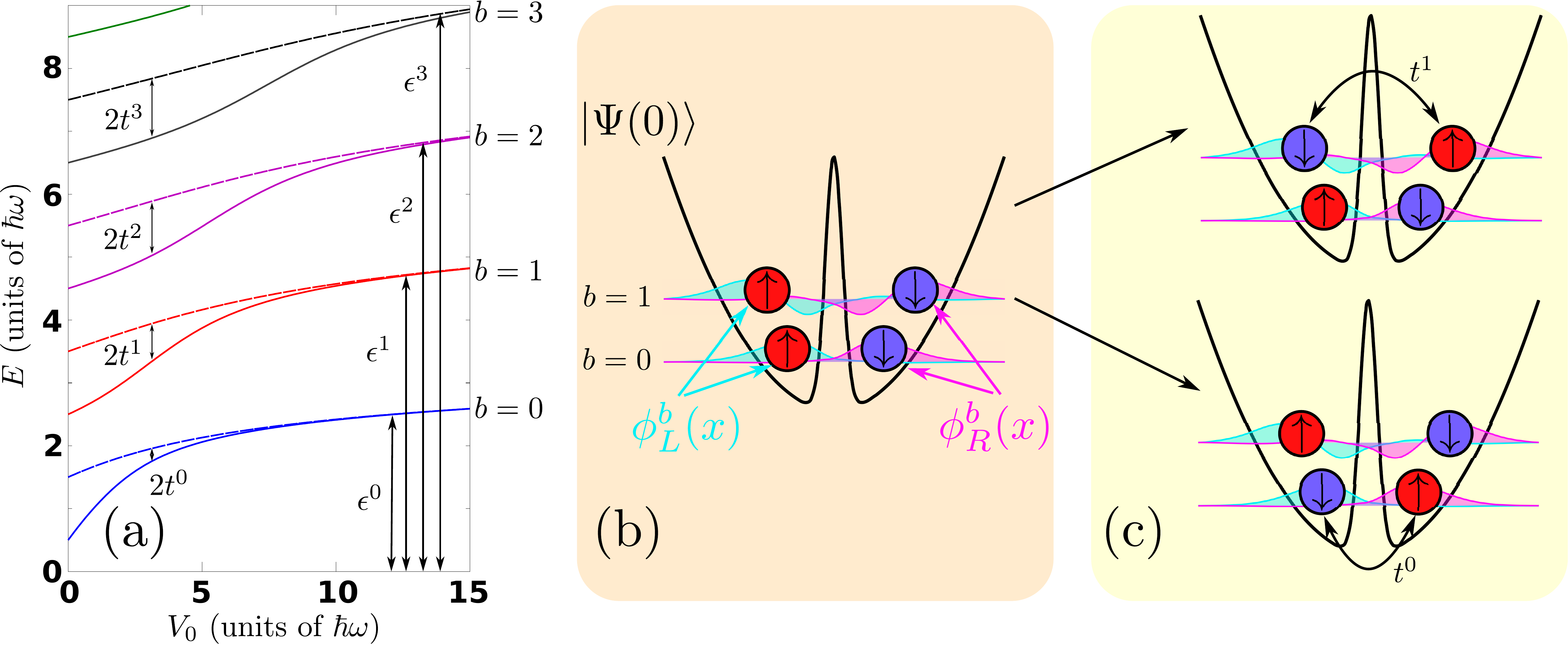}
    \caption{(a) Single-particle eigenspectrum of a DW with $w=0.5$ for a varying barrier height
        $V_0$. $b$ refers to the band index and $\epsilon^b$, $t^b$ to the energy and tunneling
        rate of the $b$ band respectively.  Schematic illustration (b) of the initial state $| \Psi
        (0)\rangle$ and (c) the possible states accessed by single-particle tunneling for
    $N_{\uparrow}=N_{\downarrow}=2$.}
    \label{fig:setup}
\end{figure}

The purpose of this work is to examine whether a phase separated state can be stabilized in the
presence of interactions and reveal its relation to the (ferro)magnetic properties of the system.
A promising candidate for such an investigation is the initial state 
\begin{equation}
    | \Psi(0) \rangle =\prod_{b=0}^{N_\uparrow-1} \underbrace{\int {\rm d}x~\phi_{L}^b(x)
\hat{\psi}^{\dagger}_{\uparrow}(x)}_{\equiv \hat{a}^{b\dagger}_{L\uparrow}} \prod_{b=0}^{N_\downarrow-1}
\underbrace{\int {\rm d}x~\phi_{R}^b(x) \hat{\psi}^{\dagger}_{\downarrow}(x)}_{\equiv \hat{a}^{b\dagger}_{R\downarrow}} | 0 \rangle,
    \label{initial_state}
\end{equation}
where $N_{\uparrow}=\frac{N}{2}$ spin-$\uparrow$ and $N_{\downarrow}=\frac{N}{2}$ spin-$\downarrow$
fermions are localized in the left and right wells respectively [see Fig. \ref{fig:setup}(b) for
$N=4$]. Recall that $\phi^b_s(x)$ denotes the Wannier state corresponding to the $s \in \{L, R\}$ well and band
$b$. Herein, we intend to address the dynamics of a system initialized in the state described by Eq.
(\ref{initial_state}), especially focussing on the stability properties of the phase separation.
Evidently, in the non-interacting case each one of the constituting particles will perform its
individual tunneling oscillation with a frequency $2 t^b$ and, consequently, the phase separation
imprinted in the initial state will be periodically lost and recovered during the time-evolution.
However, in the case of $g \neq 0$ the individual tunneling channels of each of the particles couple
due to the interparticle interaction. The interaction between the spin components is particularly
important since the system accesses via tunneling, states possessing a substantial density overlap
for anti-aligned spins yielding an interaction energy $E_I \sim g \sum_{s \in \{L,R\}} \int
    {\rm d}x | \phi^b_s(x)|^2 | \phi^{b'}_s(x)|^2$, see for instance Fig. \ref{fig:setup}(c).
    Employing a mean-field argumentation one arrives at the conclusion that the tunneling among the
    wells slows down and eventually terminates as the repulsion increases.  This is due to the large
    interaction energy of a spin-$\uparrow$ and a spin-$\downarrow$ atom occupying the same well
    when compared to the interaction energy contained in $| \Psi (0) \rangle$ where the
    spin components are phase separated.  However, the interparticle interaction possibly induces
    two- (or more) body correlations crucially affecting the dynamics of the system. As we shall
    demonstrate later on this is indeed the case and the dynamics for $g \neq 0$ is more involved
    than what is expected by the above-mentioned mean-field argumentation.

\subsection{Magnetization Imbalance and MB Eigenstate Categorization in Terms of Bands}\label{sec:hamilt_obs}
To monitor the degree of phase separation between the spin components during the dynamics of the
system we employ the experimentally accessible measure
\cite{Valtolina} 
\begin{equation}
    M = \frac{1}{2}({M}_{\uparrow}-{M}_{\downarrow}), \hspace{2pt} {\rm with} \hspace{2pt} 
    {M}_{\alpha}=\frac{1}{N_{\alpha}}\left(\int^0_{-\infty} dx~\rho^{(1)}_{\alpha}(x;t) - \int_0^\infty
dx~\rho^{(1)}_{\alpha}(x;t)\right).
\label{magn}
\end{equation}
Here $\rho^{(1)}_{\alpha}(x;t) = \langle \Psi(t) | \hat{\psi}^\dagger_{\alpha}(x)
\hat{\psi}_{\alpha}(x) | \Psi(t) \rangle$ is the spin-dependent, $\alpha \in \{ \uparrow,
\downarrow\}$, one-body density. Notice that both the Hamiltonian, Eq. (1), and the initial state, Eq. (3),
    are invariant under the transformation $x \to -x$, $| \uparrow \rangle \to | \downarrow \rangle $ and $| \downarrow \rangle  \to | \uparrow \rangle$, implying that $M_{\uparrow}+M_{\downarrow}=0$ is conserved during the dynamics. The quantity $M$ takes its extreme values $M=1$ and
$M=-1$ when the particles within each of the wells are fully-polarized, a situation equivalent to
a perfect phase separation. The sign of $M$ in this case depends on whether the spin-$\uparrow$
particles reside in the left ($M=1$), as is the case for $| \Psi(0) \rangle$, or right ($M=-1$)
well. In the case that $M=0$ the spin-$\uparrow$ and spin-$\downarrow$ particles are distributed over
both wells showing that the spin components are miscible. Since $M \neq 0$ corresponds to states
magnetized along the $x$ spatial-axis [see also Eq.(3)], $M$ will be herewith referred to as magnetization imbalance.

Furthermore, let us note that for large barrier heights and weak or intermediate interactions, we
expect that the band-gaps between the non-interacting bands constitute the largest energy scale of
the system, see Fig. \ref{fig:setup}(a).  As a consequence, the energetic characterization of the MB
eigenstates in terms of non-interacting bands will be of great importance in the following. We
assign each eigenstate of the non-interacting $N$-body system, $| \Psi_{g=0} \rangle$ to an
energetic class by employing the vector $\vec{n}_B=(n^0_B,n^1_B,\dots)$. This vector contains the
occupation numbers of each of the non-interacting bands,  $n^{b}_B=\langle \Psi_{g=0} | \hat
n^{b}_{L\uparrow}+\hat n^{b}_{R\uparrow}+\hat n^{b}_{L\downarrow}+\hat n^{b}_{R\downarrow} |
\Psi_{g=0} \rangle$ ($0 \le n^b_B \le 4$), with $\hat n^{b}_{s \alpha}$ being the number operator
that counts the number of spin-$\alpha$ particles residing in the Wannier state $\phi_s^{b}(x)$.
Accordingly, each eigenstate of the interacting system, $g \neq 0$, will be assigned to an energy
class, $\vec{n}_B$, if it constitutes a superposition of non-interacting eigenstates of this
particular class. For instance the initial state, $| \Psi (0) \rangle$, belongs to the
$\vec{n}_B=(2,2,0,\dots)$ class for $N=4$, see also Fig. \ref{fig:setup}(b). Indeed, the initial
state for $N=4$ contains two fermions in the $0$th band ($n_B^0=2$) and two additional ones residing
in the $1$st excited band ($n_B^1=2$).

\section{Many-Body Eigenspectrum and Correlated Dynamics}\label{sec:deep}

In this section we examine the eigenspectrum of the full MB Hamiltonian $\hat H$ [see Eq.
(\ref{hamilt})] in the case of $N=4$ fermions. Then we analyze the correlated dynamics of such a
system initialized in the state $|\Psi (0) \rangle$ [Eq. (\ref{initial_state})] and subsequently
left to evolve within $\hat{H}$. This investigation permits us to identify the emergent phase
separation behavior between the spin components for varying interaction strength. To track the
correlated dynamics of this system we employ ML-MCTDHX \cite{MLX} and, in particular, its reduction
for spin-$1/2$ fermions (for more details see Appendix C). ML-MCTDHX is an {\it ab initio}
variational method that takes all correlations into account enabling us to reveal their influence
into the static properties and in particular the dynamics of MB systems.
We generalize our results to the $N>4$ case in Appendix A.

\begin{figure}[t]
    \includegraphics[width=1.0\textwidth]{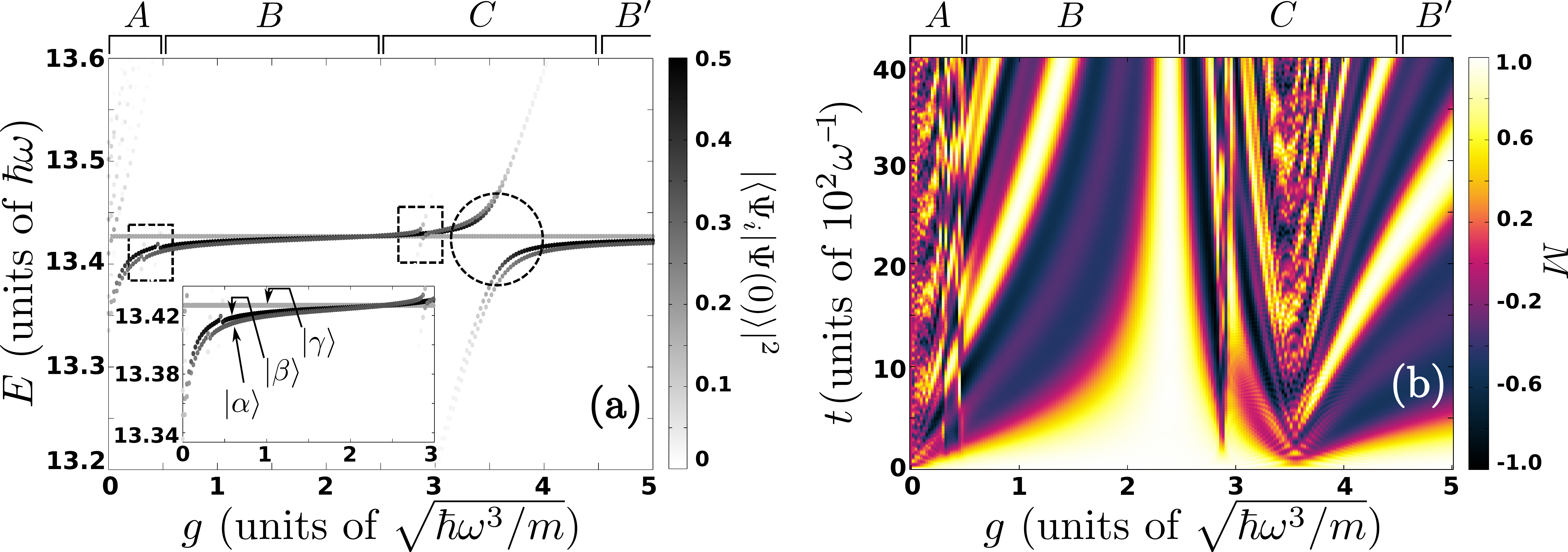}
    \caption{(a) Eigenspectrum of $N_{\uparrow}=N_{\downarrow}=2$ fermions confined in a DW for
        varying interaction, $g$. The overlap of the depicted eigenstates with the initial state $| \langle
        \Psi_i|\Psi(0)\rangle|^2$ is indicated by the color of the datapoints and satisfies the criterion $| \langle
    \Psi_i|\Psi(0)\rangle|^2>10^{-3}$. The dashed circle and boxes indicate the locations of
    the wide and narrow
    avoided crossings respectively. The inset provides a magnification
    of the eigenspectrum for $0<g<3$.  (b) Dynamics of the magnetization imbalance, $M$, for varying
$g$. In all cases $w=0.5$, $V_0=8$ and $N_\uparrow=N_\downarrow=2$. }
    \label{fig:MB_res1}
\end{figure}

\subsection{Many-Body Eigenspectrum}

The eigenspectrum of $\hat H$ [Eq. (\ref{hamilt})], for $N_\uparrow=N_\downarrow=2$ fermions and
varying $g$, is presented in Fig. \ref{fig:MB_res1}(a), in the case of a relatively deep ($V_0=8$,
$w=0.5$) DW potential.  The overlap of the MB interacting eigenstates, $|\Psi_i\rangle$ with the
initial state, $|\Psi(0)\rangle$ is indicated by the different colours in Fig.
\ref{fig:MB_res1}(a).  Based on the eigenspectrum we can identify four different interaction
regimes, indicated by $A$, $B$, $C$ and $B'$ in Fig.  \ref{fig:MB_res1}(a), where the overlap of the
initial state, $|\Psi(0)\rangle$ with the MB eigenstates $|\Psi_i\rangle$ of $\hat H$ exhibits a
qualitatively different behaviour. In addition, by expanding each eigenstate  $| \Psi_i \rangle$ in
the number states of the Wannier basis, $\phi_s^b(x;t)$, (not shown here for brevity) we are able to
identify its energetic class, $\vec{n}_B$ (see section \ref{sec:hamilt_obs}) which is important for
identifying the interband processes emanating in the eigenspectrum and dynamics.

For weak interactions, $g<0.5$ within the interaction regime $A$ we observe that multiple
eigenstates [the ones with $E>13.42$ are hardly visible in Fig. \ref{fig:MB_res1}(a)] contribute to
the initial state. We remark that these states belong to the energy class $\vec{n}_B=(2,2,0,\dots)$
according to the energy categorization given in section \ref{sec:hamilt_obs}. The energies of the
eigenstates with $E>13.42$ increases for increasing $g$, while their overlap with the initial
state decreases, see Fig.  \ref{fig:MB_res1}(a) for $0<g<0.5$. For $g \approx 0.5$ only three of the
aforementioned eigenstates with $E<13.43$ possess a significant overlap with $|\Psi (0) \rangle$ see
also the inset of Fig.  \ref{fig:MB_res1}(a). Additionally, narrow avoided crossings [see the dashed
box in Fig.  \ref{fig:MB_res1}(a) for $g \approx 0.2$] emerge but overall the MB eigenspectrum is
only slightly modified.  These narrow avoided crossings result from the coupling of states belonging
to the energy classes $\vec{n}_B=(3,0,1,0,\dots)$ and $\vec n_B=(2,2,0,\dots)$ by a weak
two-particle interband transfer process.  

Entering the interaction regime $B$, $0.5<g<2.5$, we observe that the three eigenstates of $\hat H$ possessing the dominant overlap with $| \Psi (0)
\rangle$, are quasi-degenerate.  In terms of increasing energetic order we refer to
these eigenstates as $| \alpha \rangle$, $| \beta \rangle$ and $| \gamma \rangle$, see also the
inset of Fig. \ref{fig:MB_res1}(a).  The existence of the quasi-degenerate predominantly occupied
eigenstates within the $B$ and also $B'$ ($4.5 \le g < 5$) interaction regimes implies that the
time-scales of the dynamical evolution, which are associated with the energy differences of these
quasi-degenerate states, are rather large. Therefore, these interaction regimes are very promising
for studying the dynamical stability of the phase separation exhibited by the initial state $| \Psi
(0) \rangle$. Note that the physical reasoning behind the emergence of this quasi-degenerate eigenstate manifold will be the main focus of section IV. At $g \approx 3.5$ the three aforementioned quasi-degenerate eigenstates show a wide
avoided crossing [see the dashed circle in Fig. 2(a)] with the eigenstates of the $\vec{n}_B=(3,1,0,\dots)$ energy class within the
interaction regime $C$, $2.5 < g <4$. As we shall explicate later on, this interband avoided
crossing is the fermionic analogue of the so-called cradle mode that has been identified in the
interaction quench dynamics of spinless lattice trapped bosonic ensembles
\cite{Mistakidis1,Mistakidis2,Mistakidis3}. For larger repulsions, $g>4.5$, the quasi-degeneracy of
the predominantly occupied eigenstates reappears giving rise to the $B'$ interaction regime. The
eigenspectrum for these interactions ($g>4.5$) possesses a similar structure to the one observed
within the interaction regime $B$. Note that the Tonks-Girardeau limit of our system is approached
for $g>5$ (not shown here for brevity). It is known \cite{Giann2} that in this case the
eigenspectrum features an avoided crossing between the aforementioned quasi-degenerate states and
the ones belonging to the energetically lowest class $\vec{n}_B=(4,0,\dots)$. The eigenspectrum in
this case can be theoretically described by using standard spin-chain techniques \cite{Giann2}. 
We remark that the state $| \gamma \rangle$ possesses an interaction independent
eigenenergy, associated with its fully antisymmetric character under particle exchange.

\begin{figure}[t]
    \includegraphics[width=1.0\textwidth]{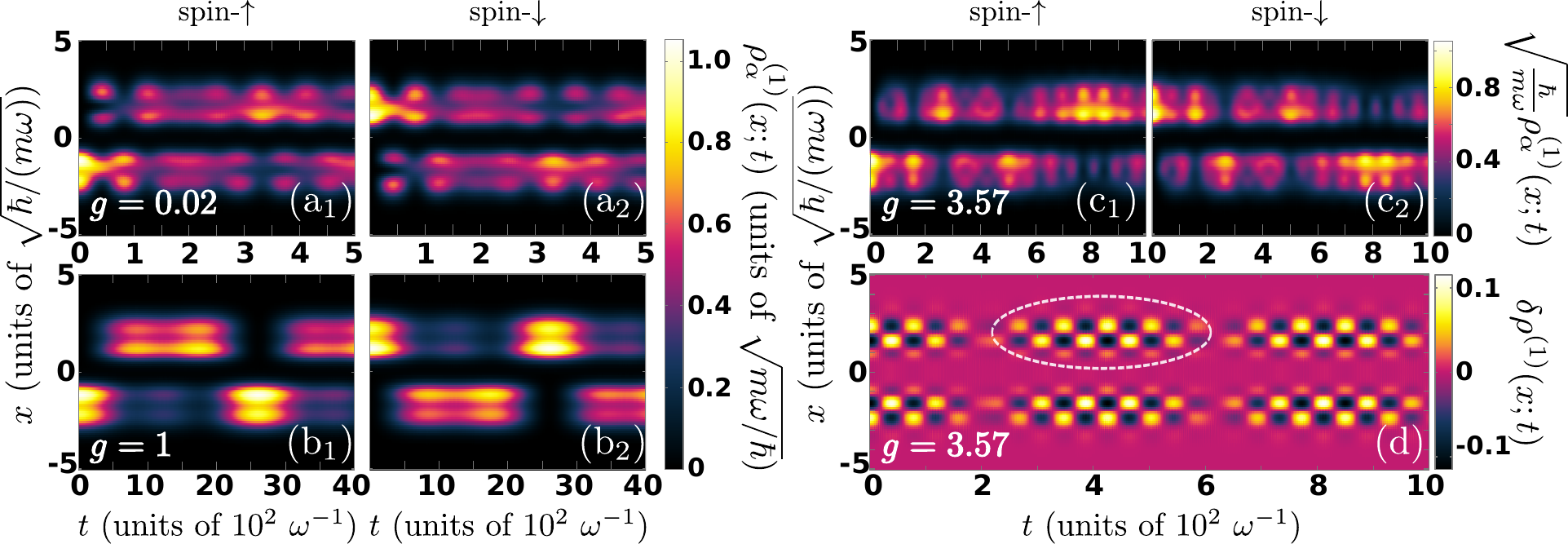}
    \caption{(a$_i$), (b$_i$), (c$_i$)
        Time-evolution of the one-body densities $\rho^{(1)}_{\alpha}(x;t)$ for
        ($i=1$) the spin-$\uparrow$ and ($i=2$) the spin-$\downarrow$ component for
        three different interaction strengths (see legends). (d) The
        total one-body density fluctuations,
        $\delta\rho^{(1)}(x;t)$ for $g=3.57$, where the cradle mode
        is clearly imprinted (see the dashed ellipse). In all cases $w=0.5$, $V_0=8$ and
    $N_\uparrow=N_\downarrow=2$. }
    \label{fig:MB_res2}
\end{figure}

\subsection{Correlated Dynamics}

To inspect the stability of the phase separation encoded in $| \Psi (0) \rangle$ for different
interaction strengths we let the system initialized in the state $| \Psi (0) \rangle$ of Eq.
(\ref{initial_state}) to evolve in time and subsequently we track the magnetization imbalance, $M$
[Eq. (\ref{magn})], during the out-of-equilibrium dynamics.  To provide a further interpretation  of
the participating dynamical modes we relate our findings regarding the phase separation to the
eigenspectrum of $\hat H$ [Eq. (\ref{hamilt})], see also Fig. \ref{fig:MB_res1}(a).  For weak
interactions, i.e. $0<g<0.5$, the phase separated, $| \Psi(0) \rangle$, state is shown to be
unstable especially when the non-interacting limit is approached. Indeed, as already identified in
the eigenspectrum [see Fig.  \ref{fig:MB_res1}(a) for the interation regime $A$] multiple
eigenstates contribute to the MB dynamics whose occupation results in a fast decay of the phase
separation. This is indeed justified by inspecting the corresponding time-evolution of $M$ showing
fast oscillations of small amplitude [Fig.  \ref{fig:MB_res1}(b)]. The dynamics is better captured
by the evolution of the spin-$\alpha$ one-body density, $\rho^{(1)}_{\alpha}(x;t)$, see Fig.
\ref{fig:MB_res2}(a$_1$) and \ref{fig:MB_res2}(a$_2$).  Here, the dominant process is single
particle tunneling. More precisely, the particles occupying the first excited band tunnel between
the wells with much higher frequency than the particles occupying the lowest band. This can be
identified by comparing the rate of tunneling of the two humped density structure ($b=1$ band)
appearing in Fig.  \ref{fig:MB_res2}(a$_1$) with the tunneling of the density residing near the
center of the well ($b=0$ band) within the time-interval $0 <t < 100$.  Additionally, an interaction
induced dephasing, due to the involvement of the multitude of eigenstates identified in Fig.
\ref{fig:MB_res1}(a), is evident as $| \Psi(0) \rangle$ does not completely revive during the time
evolution.

Further monitoring the dynamical evolution of the system we  observe that the phase separated state,
$| \Psi(0) \rangle$ [Eq. (\ref{initial_state})], is a long-lived metastable state within the
interaction regime $B$. Indeed, for $0.5<g<2.5$ we can infer the decay of the phase separation,
imprinted in the magnetization imbalance, $M$ and its subsequent revival [Fig.
\ref{fig:MB_res1}(b)]. This process is relatively fast for weak interactions within the interaction
regime $B$.  For instance, the phase separated state $| \Psi (0)\rangle$ decays to a miscible state
with $M=0$ at $t\approx 500$ for $g=1$.  Qualitatively similar dynamics occurs but it is shown to be
significantly slower for $1<g<4$, e.g. at $g=2$, $M=0$ is reached for $t\approx 2000$, while the
life-time of $|\Psi(0)\rangle$ exceeds $t=4000$ for $g \approx 2.5$ [Fig.  \ref{fig:MB_res1}(b)].
The metastability of $|\Psi(0)\rangle$ is accordingly well-justified since its life-times are much
larger than the inverse of the characteristic tunneling rate of the ground, $\pi/(2 t^0) \approx
42$, and the first excited band, $\pi/(2 t^1) \approx 190$. To shed light into the dynamical
evolution of $| \Psi(0) \rangle$ we also inspect the one-body densities of the spin components,
$\rho^{(1)}_{\alpha}(x;t)$ [Fig. \ref{fig:MB_res2}(b$_1$) and \ref{fig:MB_res2}(b$_2$)] at $g=1$.
Indeed, at $t \approx 500$ $\rho^{(1)}_{\uparrow}(x;t)$ and $\rho^{(1)}_{\downarrow}(x;t)$ are
delocalized over both wells and they are almost perfectly overlapping which is in accordance to the
value $M=0$ [see Fig. \ref{fig:MB_res1}(b)]. Note here that the absence of any signature of phase
separation within each of the wells justifies the use of $M$ as a measure of phase separation.
Subsequently, the density of each component accumulates in the opposite well, than it was residing
initially, but a small density portion remains in the initially populated well. Finally, at $t
\approx 2700$ an almost perfect revival of $| \Psi(0) \rangle$ occurs. For larger evolution times,
the above-mentioned dynamics is repeated in a periodic manner.  Regarding the underlying tunneling
mechanisms, the evolution of $\rho^{(1)}_{\alpha}(x;t)$ is indicative of a low-frequency two-body
correlated tunneling dynamics for both spin components, as the entire density of two spin-aligned
fermions seems to tunnel among the wells without being deformed. In addition, a contribution
stemming from a single-particle tunneling process is also visible in Fig.  \ref{fig:MB_res2}(b$_1$)
and \ref{fig:MB_res2}(b$_2$), notice for instance the dynamics of the faint two-humped structure for
$t\approx 600$, $t\approx 1200$ and $t=1800$.  In the following section it will be shown that the
occurrence of the interaction regime $B$ can be explained by examining the spin-order exhibited in
the system.

For strong interactions, $g>2.5$, the eigenstates belonging to the energy class
$\vec{n}_B=(3,1,0,\dots)$ cross with the predominantly occupied eigenstates of the class
$\vec{n}_B=(2,2,0,\dots)$ as shown in Fig. \ref{fig:MB_res1}(a) at $g \approx 3.5$. The states of
the two energy classes exhibit two avoided crossings (indicated in Fig. 2(a) by the dashed circle) due to the interband interaction-induced
coupling which is a manifestation of the cradle mode \cite{Mistakidis1,Mistakidis2,Mistakidis3}.
This resonant behaviour is directly imprinted on $M$, which shows a strong dependence of the
lifetime of $|\Psi(0)\rangle$ on the value of $g$, see Fig. \ref{fig:MB_res1}(b) at $g\approx 2.8$
and $g \approx 3.5$. The spin-dependent one-body densities also show a tunneling behavior similar to the weakly
interacting case, compare Fig. \ref{fig:MB_res2}(c$_1$) and \ref{fig:MB_res2}(a$_1$). The cradle mode
is manifested as a dipole-like oscillation within each well.  To explicitly demonstrate its
existence we invoke the total one-body density fluctuations
\cite{Mistakidis1,Mistakidis2,Mistakidis3} defined as 
\begin{equation} 
    \delta \rho^{(1)}(x;t)=
    \sum_{\alpha \in \{\uparrow,\downarrow\}} \left[ \rho^{(1)}_{\alpha}(x;t) -\frac{1}{T}
\int_0^{T} dt'~\rho^{(1)}_{\alpha}(x;t') \right].  
\label{eq:dens_fluct} 
\end{equation}
Indeed, $\delta \rho^{(1)}(x;t)$ reveals dipole-like oscillations within both wells [see for
instance Fig.  \ref{fig:MB_res2}(d) around $t \approx 400$ i.e. the encircled region] and a beating dynamics for the intensity
of the cradle mode. This beating can be understood by inspecting the eigenspectrum of the system
[Fig.  \ref{fig:MB_res1}(a)], where two almost perfectly overlapping cradle resonances can be
identified at $g \approx 3.5$, yielding two cradle frequencies of comparable magnitude.  Notice that
the cradle mode exhibited in our system is slightly different from its bosonic counterpart
\cite{Mistakidis1, Mistakidis2,Mistakidis3} as it does not involve overbarrier transport between the
different wells but rather a direct interband population transfer within a particular well.  The
absence of overbarrier transport can be identified in Fig. \ref{fig:MB_res2}(d) as the density
fluctuations in the spatial region of the barrier, $x \approx 0$ are vanishing.
 
\section{Interpretation of the magnetic properties and the effective tight-binding model} \label{sec:tju}

Having appreciated the magnetic properties of the system within the fully-correlated ML-MCTDHX approach,
we next proceed by constructing a reduced effective model. This model as we shall discuss below facilitates the
qualitative interpretation of the correlated MB dynamics. In particular, the qualitative
understanding of the underlying magnetic properties of the system via the effective model enables the identification of the decay
mechanisms of the phase separation in a straightforward and intuitive way, allowing also, for
comparisons with previous studies.

\subsection{The Effective Tight-Binding Model}

As already mentioned in section II, the band-gaps constitute the largest energy-scale of the system
for both weak and intermediate interactions. It is therefore, well-justified to assume that a
corresponding tight-binding model might sufficiently capture the observed dynamics.  Within such a
tight-binding model the Wannier states, $\phi_{s}^b(x)$, with $s \in \{L,R\}$, constitute the basis
states of the MB Hamiltonian. The non-interacting Hamiltonian reads $\hat H_0=-\sum_{b=0}^{\infty}
\sum_{\alpha \in \{\uparrow,\downarrow\}} t^b \left( \hat{a}^{b\dagger}_{R\alpha}
\hat{a}^b_{L\alpha} +\hat{a}^{b\dagger}_{L\alpha} \hat{a}^b_{R\alpha} \right) +\sum_{b=0}^{\infty}
\sum_{\alpha \in \{\uparrow,\downarrow\}} \epsilon^b \left(\hat{n}^{b}_{L\alpha}
+\hat{n}^{b}_{R\alpha} \right)$, where $\epsilon^b$ is the average energy of the non-interacting
eigenstates forming the band, $b$. Also, $\hat{a}^{b\dagger}_{s\alpha}$ ($\hat{a}^{b}_{s\alpha}$) is
the operator that creates (annihilates) a spin-$\alpha$ particle in the Wannier state
$\phi^b_{s}(x)$ and $\hat{n}^{b}_{s\alpha} \equiv
\hat{a}^{b\dagger}_{s\alpha}\hat{a}^{b}_{s\alpha}$. The exact form of the interaction term, $\hat
H_I$, involves all matrix elements between the different Wannier states and it is, thus, quite  
complicated in appearance.  Within the lowest-band approximation the Fermi-Hubbard model circumvents this issue by
considering only on-site interactions and neglecting all density-induced tunneling effects
\cite{Hubbardorig}. It constitutes a valid approximation for large $V_0$, where the underlying
Wannier basis-states are well-localized to the corresponding wells. Additionally, $g$ should define
a sufficiently smaller energy scale than the band gap, ensuring that no significant
interaction-induced interband tunneling, such as the cradle mode, occurs. Fermi-Hubbard models have
been very successful in describing various effects emanating in a variety of settings where DW or
lattice potentials are involved \cite{HubbardReview1,HubbardReview2}.

Therefore, it is tempting to
approximate the exact interaction term, $\hat H_I$, by the following effective one
\begin{equation}
    \hat{H}_I^{\rm dir}=g \left[ \sum_{b=0}^{\infty} U^b \left( \hat{n}^{b}_{L\uparrow} \hat{n}^{b}_{L\downarrow} +\hat{n}^{b}_{R\uparrow} \hat{n}^{b}_{R\downarrow}  \right)
                       + \sum_{b \neq b'\in [0, \infty)} J^{bb'} \left( \hat{n}^{b}_{L\uparrow} \hat{n}^{b'}_{L\downarrow} +\hat{n}^{b}_{R\uparrow} \hat{n}^{b'}_{R\downarrow} \right) \right],
                       \label{eff_int}
\end{equation}
where $J^{bb'}=\int {\rm d}x~|\phi_{L}^{b}(x)|^2|\phi_{L}^{b'}(x)|^2=\int {\rm
d}x~|\phi_{R}^{b}(x)|^2|\phi_{R}^{b'}(x)|^2$ and $U^b=J^{bb}$ refer to the inter and intraband
on-site interactions respectively. However, as it can be easily verified the last term of Eq.
(\ref{eff_int}) breaks the SU(2) symmetry of $\hat{H}$ [Eq. (\ref{hamilt})], since it does not
commute with $\hat{S}^2$. In order to avoid this artificial symmetry breaking one needs, also, to
include into the effective Hamiltonian the term
\begin{equation}
\hat{H}_I^{\rm exc}= -g \sum_{b \neq b'\in [0, \infty)}
J^{bb'} \left( \hat{a}^{b\dagger}_{L\uparrow} \hat{a}^{b'\dagger}_{L\downarrow} \hat{a}^{b'}_{L\uparrow} \hat{a}^{b}_{L\downarrow}+
\hat{a}^{b\dagger}_{R\uparrow} \hat{a}^{b'\dagger}_{R\downarrow} \hat{a}^{b'}_{R\uparrow} \hat{a}^{b}_{R\downarrow}+
\hat{a}^{b\dagger}_{L\uparrow} \hat{a}^{b'\dagger}_{L\downarrow} \hat{a}^{b'}_{L\uparrow} \hat{a}^{b}_{L\downarrow}+
\hat{a}^{b\dagger}_{R\uparrow} \hat{a}^{b'\dagger}_{R\downarrow} \hat{a}^{b'}_{R\uparrow} \hat{a}^{b}_{R\downarrow} \right).
\label{exchange}
\end{equation}
The term $\hat{H}_I^{\rm exc}$, which is present in the exact $\hat H_I$ of Eq. (\ref{hamilt}),
incorporates the effect where two fermions in different bands but on the same well can exchange
their spin due to their mutual interaction. Models that extend the Hubbard model in a similar
    manner to the above-mentioned have been employed in the context of the metal-insulator transition
emanating in $d$-electron systems, for a review see \cite{imada}.

Including all of the above-mentioned terms into an effective tight-binding Hamiltonian results in
the following multi-band tJU model
\begin{equation}
    \begin{split}
        \hat{H}_{\rm eff}= &-\sum_{b=0}^{\infty}\sum_{\alpha \in \{ \uparrow \downarrow\}} t^b \left( \hat{a}^{b\dagger}_{R\alpha} \hat{a}^b_{L\alpha}
    +\hat{a}^{b\dagger}_{L\alpha} \hat{a}^b_{R\alpha} \right) 
    +g\sum_{b=0}^{\infty} U^{b} \left( \hat{n}^{b}_{L\uparrow} \hat{n}^{b}_{L\downarrow} +\hat{n}^{b}_{R\uparrow} \hat{n}^{b}_{R\downarrow}  \right)\\
    &-g\sum_{b \neq b' \in [0, \infty)} J^{bb'} \left[\hat{\bm{S}}_{L}^{b}\cdot\hat{\bm{S}}_{L}^{b'}
        +\hat{\bm{S}}_{R}^{b}\cdot\hat{\bm{S}}_{R}^{b'} -\frac{1}{4} \left( \hat{n}^{b}_{L} \hat{n}^{b'}_{L} + \hat{n}^{b}_{R} \hat{n}^{b'}_{R}   \right) \right]\\
    &+\sum_{b=0}^{\infty}\sum_{\alpha \in \{ \uparrow \downarrow\}} \epsilon^b \left(\hat{n}^{b}_{L\alpha} +\hat{n}^{b}_{R\alpha} \right),
    \end{split}
   \label{eq:tJU_model}
\end{equation}
where $\hat{\bm{S}}^b_{s}=\hat{S}_{x;s}^b \bm{i} + \hat{S}_{y;s}^b \bm{j} +\hat{S}_{z;s}^b \bm{k}$
with $\hat{S}^b_{k;s}=\frac{1}{2}\sum_{\alpha,\beta} \sigma^k_{\alpha \beta} \hat{a}^{b \dagger}_{s
\alpha}  \hat{a}^{b}_{s \beta}$, $k \in \{x,y,z\}$, $s \in \{L,R\}$ and $\bm{i}$, $\bm{j}$, $\bm{k}$
refer to the unit vectors in spin-space and
$\hat{n}_s^b=\hat{n}_{s\uparrow}^b+\hat{n}_{s\downarrow}^b$.  $tJ$ models, where the on-site
interaction term vanishes as double site occupations are adiabatically eliminated, have been
originally employed to describe magnetic phenomena in condensed matter physics
\cite{Spalek1a,Spalek1b,Spalek1c} and later for the interpretation of some aspects of
superconductivity \cite{Spalek2,Gros,tJreview}.  Physically, the effective Hamiltonian of Eq.
(\ref{eq:tJU_model}) describes a collection of Hubbard-dimers for each band, $b$, that are coupled
by ferromagnetic (in the repulsive case $g>0$) on-site exchange interaction [second line of Eq.
(\ref{eq:tJU_model})] and are off-setted by the corresponding band energy [third line of Eq.
(\ref{eq:tJU_model})]. On-site interband exchange interactions, such as those encoded in Eq.
(\ref{eq:tJU_model}), are known as Hund interactions in condensed matter physics
\cite{Hund1,Hund2,Hund3}. The tight-binding approximation is only valid for $t^b/E^b \ll 1$ or
equivalently large $V_0$. An additional limitation of the tJU model [Eq. (\ref{eq:tJU_model})] is
that $g U^b/E^b \ll 1$ allowing for the interaction-driven interband processes to be safely
neglected. Within this model states of different energy classes, $\vec{n}_B$ do not couple and as a
consequence all the elements of $\vec{n}_B$ are conserved. As we have previously established within
the full MB system (that does not possess this symmetry) such interband effects do not alter the
eigenspectrum significantly within the interaction regimes $A$ and $B$. 

Below we argue why this model leads to a {\it metastable}, phase separated, state $| \Psi (0)
\rangle$, in the case of intermediate repulsions, qualitatively explaining the magnetic order
exhibited within the interaction regime $B$. 

\begin{figure}[h]
    \centering
    \includegraphics[width=\textwidth]{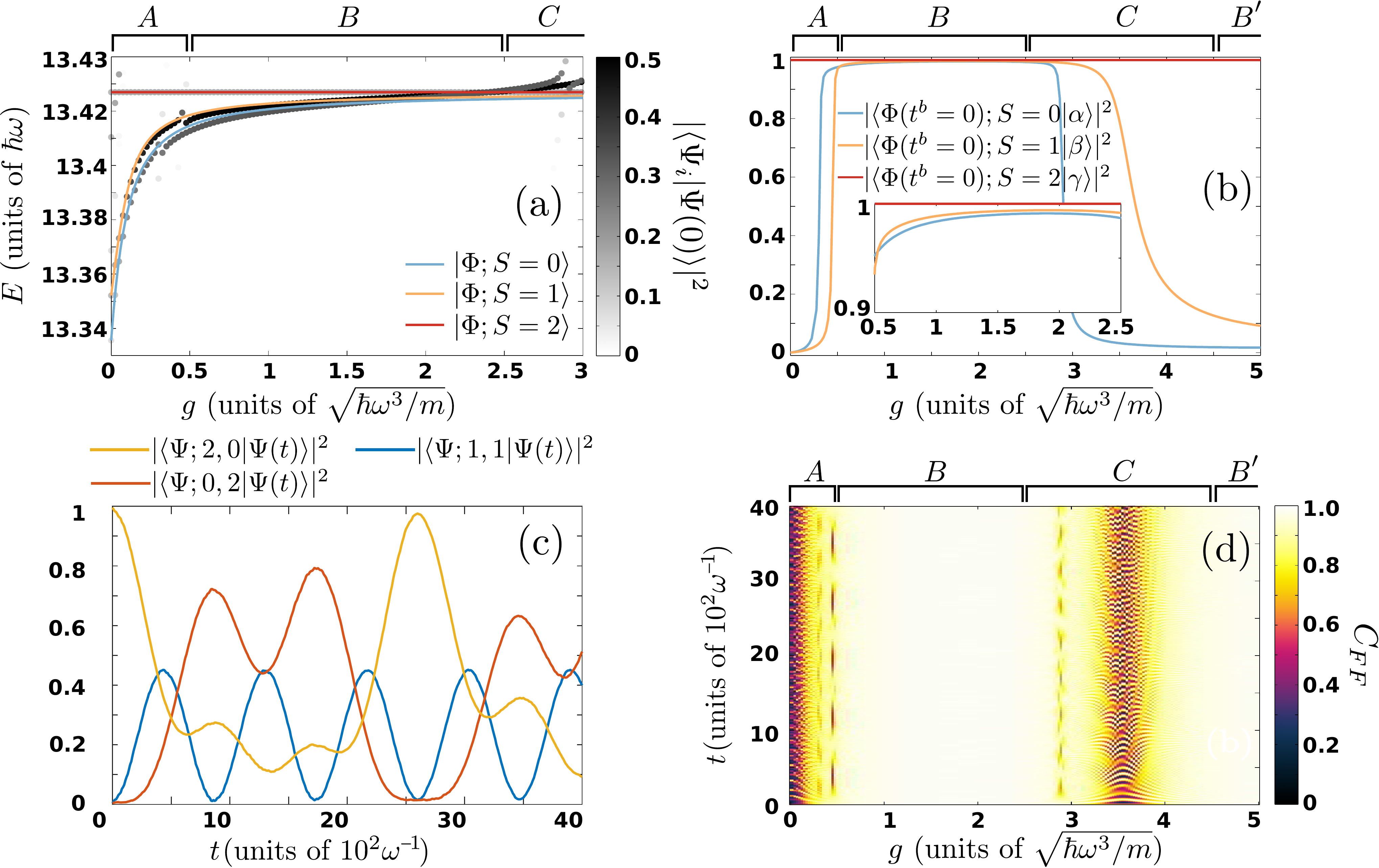}
    \caption{(a) Comparison between the eigenstates of the MB Hamiltonian, $\hat H$ (depicted by
        dots) with the eigenstates $|\Phi;S\rangle$ of the tJU model, $\hat{H}_{\rm eff}$ (colored
        lines) for varying $g$.  (b) The overlap of the MB eigenstates $|\alpha\rangle$,
        $|\beta\rangle$, $|\gamma\rangle$ with the eigenstates of the tJU model with $t^b=0$ for $b
        \in \{ 0, \dots, \frac{N}{2}-1 \}$. The inset provides a magnification of (b) within the
        interaction regime $B$, $0.5<g<2$.  (c) Time-evolution of the overlap between the MB
        wavefunction, $| \Psi(t) \rangle$, and the states $| \Psi;N_{\uparrow L},N_{\downarrow R}
        \rangle$ (see text) for $g=1$.  (d) Time-evolution of $C_{FF}$ quantifying the intra-well ferromagnetic
spin-spin correlations for varying interaction strength $g$.  In all cases $V_0=8$,
$w=0.5$ and $N_{\uparrow}=N_{\downarrow}=2$.}
    \label{fig:comparison}
\end{figure}

\subsection{Magnetic Properties of the Effective Model} \label{sec:magnprop}

Let us first discuss the relevant properties of the $N$-body eigenspectrum of the tJU model. We
operate in the $t^b/(g U^b) \ll 1$ limit, where we can neglect the tunneling term $\propto t^b$.
Indeed, for the system examined in section \ref{sec:deep} the criterion $\frac{g U^b}{t^b} \gg 1$ is
well-satisfied\footnote{For $V_0=8$ and $w=0.5$ the relevant scales for $N=4$ are $\frac{g U^1}{t^1}
\approx 11.38 g$ and $\frac{g U^0}{t^0} \approx 71.44 g$.} within the interaction regime $B$,
$0.5<g<3$.  In view of the decoupling of different energy classes $\vec{n}_B$ within the effective
tJU model we will focus on the particular energetic class that the initial state, $| \Psi(0)
\rangle$, belongs to, namely, $\vec{n}'_B$.  This class is defined as $n_B^{\prime b}=2$ for $b<N/2$
and $n_B^{\prime b}=0$ otherwise\footnote{Since we operate in the manifold of states with $S_z=0$ we
are obviously restricted to even particle numbers.}.

Focusing on the simplest case of $t^b=0$, for all involved $b$, the effective Hamiltonian can be
expanded in two intra-well Hamiltonian terms $\hat{H}_{\rm eff}=\hat{H}_R+\hat{H}_L$ that 
are decoupled. among them.  By projecting these Hamiltonian terms into the energy class $\vec{n}'_B$ the
former reads
\begin{equation}
    \hat{P}_B \hat{H}_s \hat{P}_B=
    g\sum_{b=0}^{\frac{N}{2}-1} U^{b} \hat{n}^{b}_{s\uparrow} \hat{n}^{b}_{s\downarrow}
    -g\sum_{b=0}^{\frac{N}{2}-1} \sum_{b' \neq b} J^{bb'} \left[\hat{\bm{S}}_{s}^{b}\cdot\hat{\bm{S}}_{s}^{b'}
         -\frac{1}{4}  \hat{n}^{b}_{s} \hat{n}^{b'}_{s} \right]\\
    +\sum_{b=0}^{\frac{N}{2}-1} \epsilon^b \hat{n}^{b}_{s},
    \label{spin-hamiltonian}
\end{equation}
where $\hat P_B$ is the projection operator into $\vec{n}'_B$.  Equation
(\ref{spin-hamiltonian}) corresponds to a ferromagnetic Heisenberg model, incorporating additional
energy shifts depending on the particle occupation $\propto \hat{n}_s^b$, $s \in \{ L, R\}$. For
$g>0$ the sum of these energy shifts contained within $\hat H_L$ and $\hat H_R$ is minimized in the
case that no double occupations of a particular site occur, i.e. $\langle \Psi |
\hat{n}^{b}_{s\uparrow} \hat{n}^{b}_{s\downarrow} |\Psi \rangle=0$ for all $b$ and $s$.  The spin
configuration for $t^b =0$ can be characterized by the quantum numbers $S$, $S_L$, $S_R$, where
$\hat{S}^2_s=\sum_{b,b'} \hat{\bm S}_s^b \cdot \hat{\bm S}_s^{b'}$, refers to the total spin within
the $s \in \{L,R\}$ well. It is well-known that ferromagnetic Heisenberg models exhibit
ferromagnetic ground states \cite{Mattisbook} and as a consequence the ground states of Eq.
(\ref{spin-hamiltonian}) correspond to the largest possible values of $S_L$ and $S_R$. Notice, also,
that $| \Psi(0) \rangle$ is characterized by maximum $S_L$ and $S_R$, since the spin-state within
each well is fully spin-polarized. As a consequence, we can conclude that $| \Psi(0) \rangle$ belongs
to a degenerate manifold of dimension $N/2$ at an energy $E=E_B=2 \sum_{b=0}^{N/2-1} \epsilon^b$.
This manifold consists of the states $|\Phi(t^b=0);S \rangle$ with quantum numbers
$S_L=S_R=\frac{N}{4}$ but varying total spin $S \in \{0,1,\dots,\frac{N}{2}\}$ (see also below). In
addition, the eigenstates $|\Phi(t^b=0);S \rangle$ get energetically well-separated from the other
states with $\vec{n}_B=\vec{n}'_B$ as the gap between them scales linearly with $g$, see Eq.
(\ref{spin-hamiltonian}).

The inclusion of the tunneling term for $t^b \neq 0$ induces couplings between the above-mentioned
degenerate states resulting in the lifting of their degeneracy. Indeed, by treating the tunneling term in Eq. (\ref{eq:tJU_model}) within second
order perturbation theory [see Appendix C], we can show that in the $t^{b}\ll g U^{b}$ limit the
effective Hamiltonian projected on the manifold of degenerate states spanned by $|\Phi(t^b=0);S
    \rangle$ reads
\begin{equation}
    \hat P_D \hat{H}_{\rm eff} \hat P_D = E_B + \sum_{b=0}^{\frac{N}{2}-1} \frac{4 (t^b)^2}{g
    \tilde{U}^b} \left( \hat{\bm S}^b_L \cdot \hat{\bm S}^b_R -\frac{1}{4} \right),
\label{Andexchange}
\end{equation}
where $\hat P_D$ is the projection operator $\hat{P}_D=\sum_{S=0}^{\frac{N}{2}} |\Phi(t^b=0);S
\rangle \langle \Phi(t^b=0);S |$ and the interaction parameter $\tilde{U}^b$ refers to
$\tilde{U}^b=\sum_{b=0}^{\frac{N}{2}-1} J^{b_0 b}$.  Equation (\ref{Andexchange}) provides great insight
into the structures emanating in the intra-well ferromagnetically correlated states within the
tJU model in the case of non-vanishing tunneling.  Indeed, the inclusion of tunneling for $t^b \neq
0$ results to an apparent antiferromagnetic Heisenberg exchange interaction for $g > 0$, known as
the Anderson kinetic exchange interaction \cite{Andersonkin}. Note that the total spin $\hat{
\bm{S}}=\hat{\bm{S}}_L+\hat{\bm{S}}_R$ commutes with $\hat P_D \hat{H}_{\rm eff} \hat P_D$ implying that the
eigenstates of the tJU model $| \Phi;S \rangle$ reduce within the zeroth order approximation to the
ones for $t^b=0$, i.e.  $| \Phi;S \rangle= |\Phi(t^b=0);S \rangle +\mathcal{O}(\frac{t^b}{g U^b})$.
Regarding their eigenenergies, the tJU eigenstates, $| \Phi ; S \rangle$ are expected to be
energetically ordered in terms of increasing $S$ due to the antiferromagnetic character of the
Anderson exchange interaction and be quasi-degenerate possessing energy shifts among them of the
order of $\Omega^b_{d} \sim \frac{(t^b)^2}{g U^b} \ll t^b$. 

The above properties of the tJU eigenspectrum imply that the initial state $| \Psi (0) \rangle$ being a
superposition of the eigenstates $| \Phi;S \rangle$ dephases during the time-evolution with a slow
timescale $\sim \min_b (\pi / \Omega^b_d)$.  In particular $| \Psi(0) \rangle$ can be expanded in
terms of these eigenstates by utilizing the Clebsch-Gordan coefficients leading to $\langle \Phi;S |
\Psi(0)\rangle \approx \langle \Phi(t^b=0);S |
\Psi(0)\rangle=\sqrt{\frac{(2S+1)(\frac{N}{2}!)^2}{(\frac{N}{2}-S)!(\frac{N}{2}+S+1)!}}$.  Moreover,
the maximum values of $S_L=\frac{N}{4}$ and $S_R=\frac{N}{4}$ which characterize the eigenstates $| \Phi;S
\rangle \approx |\Phi(t^b=0);S \rangle$ imply intra-well ferromagnetic correlations for particles
occupying the same well\footnote{Note that within this particular configuration in terms of
    $\vec{n}_B$ and $\vec{n}_D$, the total spin within the $s$ well solely depends on the
corresponding spin-spin correlator,
$\hat{P}\hat{S}^2_s\hat{P}=\frac{N}{2}\frac{1}{2}\left(\frac{1}{2} +1 \right) + 2 \sum_{b>b'}
\hat{\bm S}^b_s \cdot \hat{\bm S}^{b'}_s$.} and stem from the ferromagnetic Hund exchange
interactions emanating in Eq. (\ref{eq:tJU_model}).  Therefore, the emergence of the interaction
regime $B$ can be attributed to the dominant contribution of the intra-well ferromagnetic
correlations when compared to the above-mentioned effective antiferromagnetism stemming from $t^b$,
see Eq. (\ref{Andexchange}). We remark here that the nature of this effective antiferromagnetism has
been identified and studied by employing spin-chain models tailored to operate in the vicinity of
the Tonks-Girardeau limit, $g \to \infty$
\cite{Deuretzbacher,Zinnerchain,Levinsen,Yangchain,Cui3,Giann2}. Note also that this notion of
antiferromagnetism does not conflict with our notion of ferromagnetism as the first is an effective
magnetic phenomenon induced by the tunneling, $t^b$, while the second is a result of the exchange
interaction term in Eq. (\ref{exchange}).

\subsection{Comparison with ML-MCTDHX}
Before analyzing further the magnetic properties of the system and their connection to the emergent
tunneling dynamics let us first establish that the magnetic properties exhibited in the framework of
the tJU model carry forward to the fully correlated case.  To this end we shall compare the
eigenspectra obtained within the tJU model with the ML-MCTDHX method.

The relevant eigenenergies within the tJU model [Eq. (\ref{eq:tJU_model})] appear in Fig.
\ref{fig:comparison}(a) as colored lines referring to the case $N=4$. Here the three eigenstates
$|\Phi;S\rangle$, with $S=0,1,2$ possess three distinct eigenenergies at $g \approx 0$. The energy
difference between the $S=0$ and $S=1$ states is given by $t^0$ and the one between the $S=1$ and
$S=2$ corresponds to $t^1$.  This decrease of the single-particle energy of $|\Phi;S=1\rangle$ and
$|\Phi;S=0\rangle$ stems from the occurrence of one and two doublons respectively in the $g \approx
0$ case. The formation of these doublons implies the double occupation of the single-particle state
$[\phi_L^b(x)+\phi_R^b(x)]/\sqrt{2}$, with $b=0,1$.  For increasing $g$ the energy of the $S=1$ and
$S=0$ eigenstates is larger due to the involvement of these doublons which contribute a substantial
amount of interaction energy.  Most importantly, for $0.5<g<2.5$ (interaction regime $B$) the
energies of the eigenstates $|\Phi;S \rangle$ converge towards the eigenenergy of
$|\Phi;S=2\rangle$, possessing $E_{S=2}(g)=E_B \approx 13.424$, and this leads to the formation of
the quasi-degenerate manifold, identified also in Fig. \ref{fig:MB_res1}(a). It can also be checked that
the energy differences between the states $|\Phi;S\rangle$ are consistent with Eq.
(\ref{Andexchange}) possessing a characteristic energy scale of $\Omega^1_d \approx 0.033/g$.
Figure \ref{fig:comparison}(a) further reveals that the eigenstates of the tJU model follow closely
the behaviour of the eigenstates of the MB system, represented as dots in Fig.
\ref{fig:comparison}(a), within both the interaction regimes $A$ and $B$. There are a few
discrepancies associated with the avoided crossings emerging in the interaction regimes $A$ and $C$
which, as also mentioned in section III, stem from the couplings between states with different
$\vec{n}_B$. Such couplings are indeed neglected within the tJU model. Nevertheless, the agreement
within the interaction regime $B$ is almost perfect and it can be further shown that the key ingredients
of the magnetic order within the tJU model are also exhibited within the fully correlated case.
Indeed, the overlaps between the MB eigenstates $|\alpha\rangle$, $|\beta\rangle$, $|\gamma\rangle$
and the initial state, $|\Psi(0)\rangle$ agree well with those found within the effective tJU
description [see Fig.  \ref{fig:comparison}(a)], namely, $|\langle \alpha | \Psi(0) \rangle|^2
\approx |\langle \Phi;S=0 | \Psi(0) \rangle|^2 =\frac{1}{3}$, $|\langle \beta | \Psi(0) \rangle|^2
\approx |\langle \Phi;S=1 | \Psi(0) \rangle|^2 =\frac{1}{2}$ and $|\langle \gamma | \Psi(0)
\rangle|^2 \approx |\langle \Phi;S=2 | \Psi(0) \rangle|^2 =\frac{1}{6}$.
Furthermore, in Fig. \ref{fig:comparison}(b) we demonstrate the large overlap of the MB eigenstates
$| \alpha \rangle$, $| \beta \rangle$ and $| \gamma \rangle$ with the eigenstates, $| \Phi(t^b=0); S \rangle$, of the tJU model
for $t^b \ll g U^b$ within the interaction regime $B$. Indeed, this
overlap is in excess of $95\%$ [see also the inset of Fig. \ref{fig:comparison}(b)], a result which is
also consistent with the values obtained within the tJU model for the overlaps $|\langle {\Phi}; S |
\Phi(t^b=0); S \rangle|^2$ (not shown for brevity). The above mentioned findings explicitly
showcase that the magnetic order exhibited in the interaction regime $B$ within the tJU model
carries forward to the MB case.
However, for stronger interactions and as the interaction regime $C$ is approached, e.g. see 
Fig. 4(b) at $g \approx 2.5$, the overlap of the MB eigenstates and the $| \Phi(t^b=0); S
\rangle$ states decreases. This feature is beyond the tJU model description and occurs due to the
interband coupling caused by the presence of the cradle mode. 

\subsection{Relation of the Magnetic Properties to the Tunneling Dynamics}

Having identified the magnetic order of the interaction regime $B$ within the full MB approach by
comparing to an effective model, we subsequently showcase the relation of these magnetic properties
to the tunneling dynamics of the system, see also Fig.  \ref{fig:MB_res2}(b$_1$) and
\ref{fig:MB_res2}(b$_2$). To unravel this interplay we define the states with $S_L=S_R=\frac{N}{2}$
and a definite spin projection $S_{z;s}=\sum_b S^b_{z;s}$ within each of the wells, namely
\begin{equation}
    | \Psi ; N_{\uparrow L}, N_{\uparrow R} \rangle= (\hat S_{-;L} \hat
    S_{+;R})^{\frac{N}{2}-N_{\uparrow L}} | \Psi (0) \rangle.
    \label{position_states}
\end{equation}
Here $\hat S_{\pm;s}=\sum_b \hat S_{x;s}^{b} \pm i \hat S_{y;s}^{b}$ refer to the spin increasing
and lowering operators within the $s \in \{ L,R \}$ well.  Note that it can be verified that the
states $|\Psi;N_{\uparrow L}, N_{\uparrow R} \rangle$ are related to the states $| \Phi(t^b=0); S
\rangle$ by a unitary transformation. But in contrast to $| \Phi(t^b=0); S \rangle$ the states
$|\Psi;N_{\uparrow L}, N_{\uparrow R} \rangle$ have a definite number of spin-$\uparrow$ and
spin-$\downarrow$ particles within each well. The expansion of $|\Psi;N_{\uparrow L}, N_{\uparrow R}
\rangle$ in the basis $| \Phi(t^b=0); S \rangle$ can be easily obtained by employing the
Clebsch-Gordan coefficients \cite{Tanoudji}.

\begin{figure}[h]
    \centering
    \includegraphics{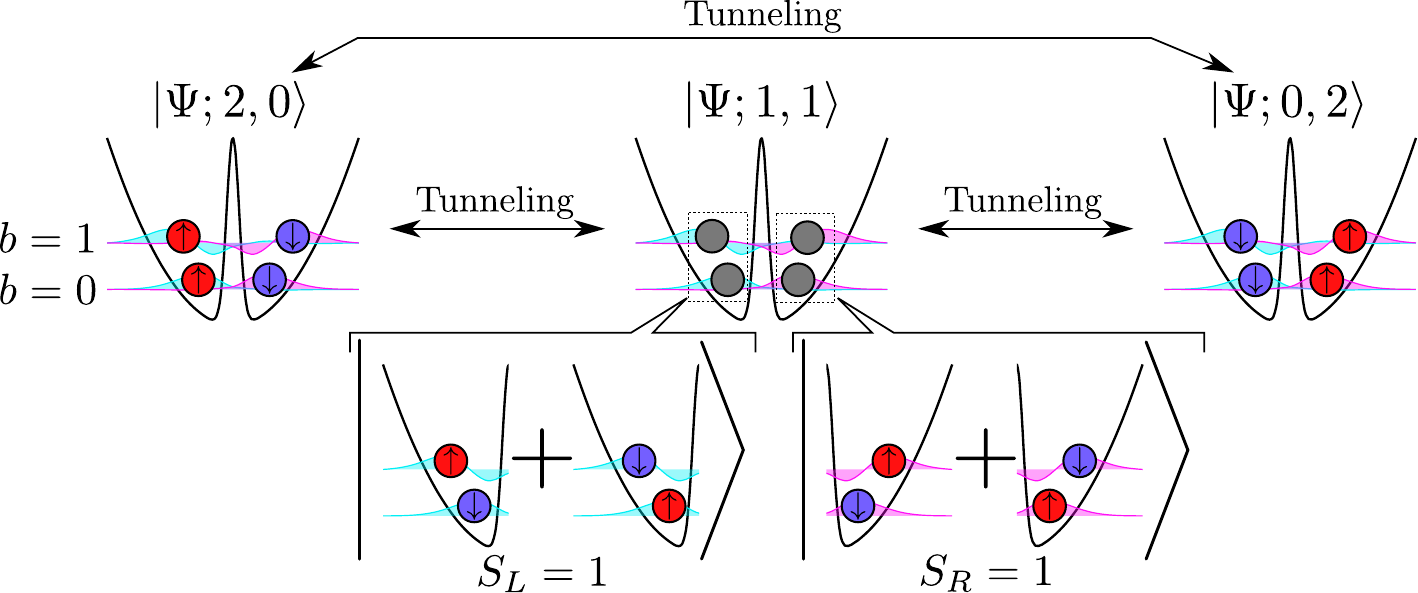}
    \caption{Schematic illustration of the correlated tunneling processes involving the
    intra-well ferromagnetically correlated states $| \Psi; N_{\uparrow L},N_{\uparrow R} \rangle$
for $N_{\uparrow}=N_{\downarrow}=2$.}
    \label{fig:schematic}
\end{figure}
The introduction of this new basis relates the phenomenon of quasidegeneracy of the states
$|\Phi(t^b=0);S\rangle$ exhibited both within the tJU model ($|\Phi;S \rangle \approx
|\Phi(t^b=0);S\rangle$) and the full MB case ($| \alpha \rangle \approx |\Phi(t^b=0);S=0\rangle$,
    $| \beta \rangle \approx |\Phi(t^b=0);S=1\rangle$ and $| \gamma \rangle \approx
|\Phi(t^b=0);S=2\rangle$), see Fig.  \ref{fig:comparison}(b), with the emergent tunneling processes.
Owing to the unitary transformation between the states $|\Psi;N_{\uparrow L}, N_{\uparrow R}
\rangle$ and the approximate eigenstates $| \Phi(t^b=0) ; S \rangle$, the accumulation of relative
phases between the eigenstates during the dynamics [due to their different eigenenergies, see
Fig. \ref{fig:comparison}(a) and Eq. (\ref{Andexchange})], corresponds to a population transfer between
the $| \Psi ; N_{\uparrow L}, N_{\uparrow R} \rangle$ states and hence to an apparent tunneling
dynamics within the spin components. For the particular case of $N=4$ particles this mechanism is
illustrated in Fig.  \ref{fig:schematic}. Notice that due to the strongly-correlated nature of the
involved states [see also Fig. \ref{fig:schematic}] such a mechanism is absent within the
Hartree-Fock mean-field theory since $|\Psi;N_{\uparrow L}=1, N_{\downarrow R}=1 \rangle$
cannot be written as a single Slater determinant.

To explicitly demonstrate the occurrence of this tunneling mechanism we present in Fig.
\ref{fig:comparison}(c) the overlap of the time-dependent wavefunction, $| \Psi (t) \rangle$, with
the states $| \Psi ; N_{\uparrow L}, N_{\uparrow R} \rangle$ for $g=1$ (interaction regime $B$). For
$0<t<500$ we observe a population transfer process from the initial state $|\Psi(0)\rangle = |\Psi;
2, 0\rangle$ to the states $|\Psi; 1, 1\rangle$ and $|\Psi; 0, 2\rangle$.  Recall that these two
processes have, also, been identified in the time evolution of $\rho^{(1)}_{\sigma}(x;t)$ [see Fig.
    \ref{fig:MB_res2}(b$_1$), \ref{fig:MB_res2}(b$_1$) and section \ref{sec:deep}]. Most
    importantly, the intricate relation of the tunneling dynamics to the magnetic properties of the
    system is now evident via employing the unitary transformation connecting the states $|\Psi;
    N_{\uparrow L}, N_{\uparrow R} \rangle$ to the eigenstates $| \Phi(t^b=0);S \rangle$.  For
    later times $t \approx 2600$ an almost perfect revival of the state $|\Psi; 2, 0\rangle$ is
    exhibited owing to the commensurability of the frequencies of these particle transfer
    processes. Indeed, the two-body tunneling process  $|\Psi; 2, 0\rangle \leftrightarrow |\Psi; 0,
    2\rangle$ is found to possess a roughly three times smaller frequency than the
    single-particle tunneling mode $|\Psi; 2, 0\rangle \leftrightarrow |\Psi; 1, 1\rangle$ [see
    Fig. \ref{fig:comparison}(c)].

\subsection{Spin-Spin Correlations}\label{sec:corr}

Having described in detail the interconnection of the magnetic properties of the system and its tunneling
dynamics we are able to shed light onto the relation of ferromagnetism and phase separation.  The
ferromagnetic order of a Fermi gas is characterized by the spin polarization and the spin-spin
correlations of the system. The total spin polarization and the total spin of the system, the latter
being related to the spin-spin correlator \cite{Koutent}, are constant during the dynamics due to
the symmetries of the Hamiltonian [Eq.  (\ref{hamilt})]. As a consequence no global ferromagnetic
order can appear during the dynamics due to the conservation laws stemming from the above
symmetries.  However, as the tJU model reveals the intra-well magnetic properties are important for
the adequate description of the system.  In this spirit, the quantity $M$ besides being a measure of
the phase separation also quantifies the spin polarization within each well [Eq. (\ref{magn})].  An
adequate quantity that captures the intra-well magnetic correlations is also hinted by the effective
tJU model. This refers to the total spin within each of the wells, $\langle \Psi(t) | \hat S^2_s |
\Psi(t) \rangle$, with $s \in \{L,S\}$. In particular, we can employ a more refined quantity by
involving some of the magnetic properties of the system identified within the tJU model.  As we have
previously discussed, the subset of states $| \Phi(t^b=0); S \rangle$ are characterized by
ferromagnetic spin-spin correlations within each well since $S_L=S_R=1$. Specifically, $|
\Phi(t^b=0);S \rangle$ are the only states within the configuration $\vec n_B=(2,2,0,\dots)$ that
exhibit this property (see also section IVB). It is thus instructive to evaluate the overlap of the
MB wavefunction, $|\Psi(t) \rangle$ with the states $| \Phi(t^b=0); S \rangle$, i.e.
$C_{FF}=\sum_{S=0}^2 |\langle \Phi(t^b=0); S | \Psi (t) \rangle|^2$. $C_{FF}$ is an adequate
quantity for studying the spin-spin correlation properties of the system, as it constitutes a lower
bound for the values of the intra-well spin-spin correlator $\langle \Psi(t) | \hat S^2_L | \Psi(t)
\rangle \ge 2 C_{FF}$ and $\langle \Psi(t) | \hat S^2_R | \Psi(t) \rangle \ge 2 C_{FF}$.
Accordingly, large values of $C_{FF}$ indicate that both wells are {\it simultaneously}
characterized by intra-well ferromagnetic spin-spin correlations.

The time evolution of $C_{FF}$ is presented in Fig. \ref{fig:comparison} (d) for varying interaction
strength $g$. For weak interactions, within the interaction regime $A$, $C_{FF}$ exhibits rapid
fluctuations between zero and unity manifesting the periodic decay and revival of the intra-well
ferromagnetic spin-spin correlations of the initial state, $| \Psi(0) \rangle$. Recall that the
phase separation, and hence the intra-well spin polarization, is unstable in this interaction regime
exhibiting decay and revival oscillations, see also Fig.  \ref{fig:MB_res1}(b), 
\ref{fig:MB_res2}(a$_1$) and \ref{fig:MB_res2}(a$_2$).  However, in the interaction regime $B$, we observe that the spin-spin
correlations within each well are ferromagnetic since $C_{FF}=1$. Indeed, the value of $C_{FF}$ is
almost constant and possesses a large value being of the order of $C_{FF} \approx 0.98$, see Fig.
\ref{fig:comparison}(d). The weak fluctuations of $C_{FF}$ around this average value, further,
showcase the stability of the intra-well ferromagnetic order. Note also that the intra-well
spin polarization quantified by $M$ is characterized as metastable within the interaction regime
$B$.  Entering the interaction regime $C$ ($2.5<g<4.5$) we observe that $C_{FF}$ exhibits
multi-frequency oscillations. These oscillations can be explained in terms of the observed resonance
of the cradle mode which introduces an interband coupling\footnote{Recall that the cradle mode
    involves states of the $\vec n_B=(3,1,0,\dots)$ energetic class. Due to the presence of three
    particles in the first band this number-state class cannot support states with $S_L=S_R=1$ and
its influence is detrimental to the ferromagnetic order.}, see also Fig. \ref{fig:MB_res1}(a) and
\ref{fig:MB_res2}(d).  For even stronger interactions, i.e. within the interaction regime $B'$,
the intra-well ferromagnetic order is reestablished and it is characterized by large and almost
constant values of $C_{FF}$ during the dynamics. 

The above results manifest the close relation between the intra-well ferromagnetic order emanating
in a DW trap and the ferromagnetic order emerging in a harmonic trap with weakly broken SU(2)
symmetry, as it has been demonstrated in Ref. \cite{Koutent}.  This order appears for intermediate
interactions where the ferromagnetic Hund exchange interaction, stemming from spin-exchange
interaction processes, see e.g. Eq. (\ref{exchange}), dominates and leads to largely stable
ferromagnetic spin-spin correlations but a fluctuating polarization. The different imposed potential
alters the manifestation of this ferromagnetic order during the dynamics. In the case of the DW the
ferromagnetic order is exhibited locally within each of the wells and as discussed above implies a
metastable phase separated state for the system. In contrast, in the case of a harmonic trap the
emergent ferromagnetic order affects the global values of the spin polarization and total spin 
implying miscibility of the contributing spin components \cite{Koutent}. This difference stems from
the Hund exchange interaction between two fermions which is only sizable if the involved
single particle states (i.e. the orbitals) possess a significant density overlap, see also Eq.
(\ref{exchange}). To understand this analogy further in the following section we study the dynamics
of the DW by employing an additional potential that breaks the SU(2) symmetry of the system.

\section{SU(2) violating case} \label{sec:Su2_viol}

Up to this point, we have identified the metastability of the phase separated state in a DW due to
the presence of the SU(2) invariance of the system. Also we have characterized the emerging
metastability of the phase separated state appearing for intermediate interactions and connected it
to the magnetic properties of the system. Next we aim to show that the phase separated state is
stable within region B even in the case that the SU(2) symmetry is weakly broken. Also, in analogy
to Ref. \cite{Koutent} the intra-well ferromagnetic correlations are shown to persist within this
region. Moreover, the implications regarding the magnetic order exhibited in a DW are briefly
discussed.

To study the case of a system with broken SU(2) symmetry we employ a linear gradient of the magnetic
field which shifts the energies of the spin-$\uparrow$ and spin-$\downarrow$ fermions in a
spatially-dependent manner. The corresponding term which is incorporated in the MB Hamiltonian of
Eq. (\ref{hamilt}), reads
\begin{equation}
    \hat{H}_g=B_0 \int {\rm d}x~x \hat{\psi}^{\dagger}_{\alpha}(x) \sigma^{z}_{\alpha \alpha'} \hat{\psi}_{\alpha'}(x).
    \label{eq:grad}
\end{equation}
The value of $B_0$ determines the energy offset between the two wells for the different spin
components.  A positive value of $B_0$ means that it is energetically preferable for the
spin-$\uparrow$ atoms to occupy the right-well and the spin-$\downarrow$ atoms the left-well.
Accordingly, when $B_0<0$ it is favorable for the spin-$\uparrow$ and spin-$\downarrow$ atoms to
occupy the left and the right-well respectively.

\begin{figure}[h]
    \centering
    \includegraphics[width=1.0\textwidth]{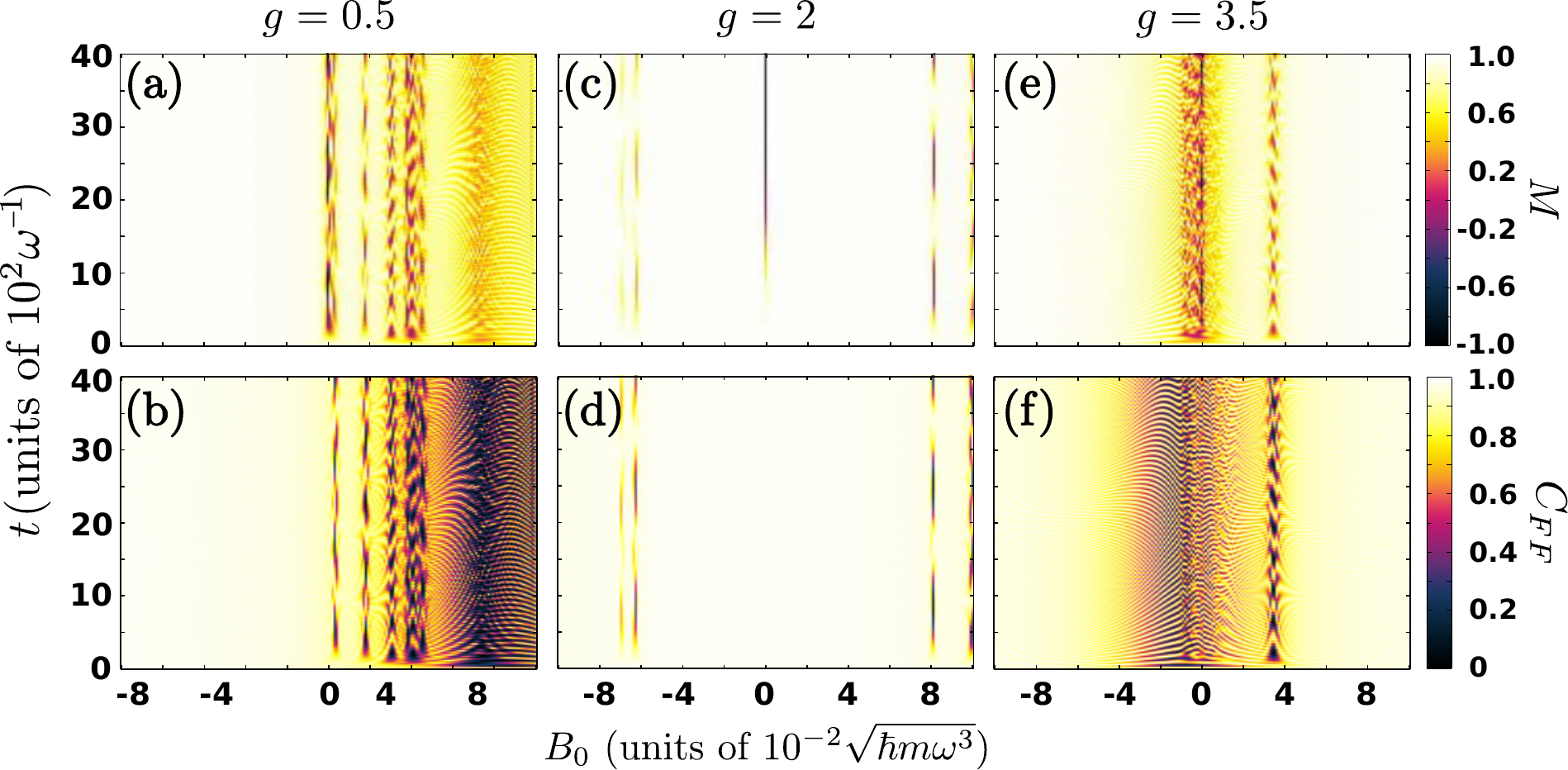}
    \caption{Time-evolution of (a), (c), (e) the magnetization imbalance, $M$, and (b), (d), (f) the
        spin-spin correlation measure $C_{FF}$ (see text) for varying strength of the linear
        gradient of the magnetic potential $B_0$.  The corresponding interaction strengths are (a),
        (b) $g=0.5$, (c), (d) $g=2$ and (e), (f) $g=3.5$. In all cases $w=0.5$, $V_0=8$ and
    $N_{\uparrow}=N_{\downarrow}=2$.}
    \label{fig:SU2viol}
\end{figure}

Figure \ref{fig:SU2viol} illustrates the time-evolution of the magnetization imbalance $M$ and
$C_{FF}$ which quantifies the degree of intra-well ferromagnetic correlations, for varying $B_0$ at
three different values of $g$ corresponding to the interaction regimes $A$, $B$ and $C$. We observe
that in the weakly-interacting case [belonging to the interaction regime $A$ in Fig.
\ref{fig:SU2viol}(a) and \ref{fig:SU2viol}(b)] and for $B_0<0$ both $M$ and $C_{FF}$ are stable
throughout the time-evolution indicating that the system remains close to its initial state. For
$B_0>0$ a multitude of resonances appear at different intervals of $B_0$ involving prominent
tunneling as captured by $M$ [Fig. \ref{fig:SU2viol}(a)].  Also, $C_{FF}$ reveals that the state of
the system is driven away from the $S_L=S_R=1$ manifold [Fig. \ref{fig:SU2viol}(b)] since
$C_{FF}<1$.  These resonances correspond to possible tunneling pathways where the spin-$\uparrow$ particles
occupying initially the left-well of the DW resonantly tunnel to the right-well (or to the opposite
direction for the spin-$\downarrow$ atoms) leading to the decay of the intra-well ferromagnetic
order.

For $g=2$ (interaction regime $B$) it can be deduced that besides the very narrow region around the
SU(2) symmetric case, i.e. at $B_0 =0$, the phase separated initial state  is stable for
$|B_0|<0.04$ as $M(t) \approx 1$ throughout the evolution [see Fig. \ref{fig:SU2viol}(c)].
Notice also that within these values of $|B_0|$ the ferromagnetic intra-well order is stable as
indicated by $C_{FF}(t) \approx 1$ [Fig. \ref{fig:SU2viol}(d)]. The stable phase separated state
appears due to the quasi-degeneracy of the states $|\alpha\rangle$, $|\beta\rangle$,
$|\gamma\rangle$ in the SU(2) preserving case for the interaction regime $B$. As stated in the
previous sections these states, owing to their intra-well ferromagnetic correlations, lie in an
energy region of the MB spectrum where no other eigenstates appear and are quasi-degenerate
characterized by a different value of the total spin $S$ [see also Fig. \ref{fig:comparison}(a)].
Recall that these states possess $C_{FF}\approx 1$ indicating their intra-well ferromagnetic
character. Moreover, their energetic ordering in terms of increasing $S$ manifests the presence of the
weak antiferromagnetic Anderson exchange interaction, [see section \ref{sec:magnprop} and Eq.
(\ref{Andexchange})]. By breaking the SU(2) symmetry with the additional spin-dependent potential
described by Eq. (\ref{eq:grad}) the states $| \alpha \rangle$, $| \beta \rangle$ and $| \gamma
\rangle$ couple with one another resulting in the formation of eigenstates with definite number of
spin-$\uparrow$ and spin-$\downarrow$ atoms in each of the wells (results not shown here for
brevity).
Therefore, for decreasing $B_0<0$ the initial state, $|\Psi(0)\rangle$, becomes the lowest-in-energy
state with $S_L=S_R=1$, while it corresponds to the highest-in-energy eigenstate of the same
manifold of states for $B_0>0$.  In both cases the phase separation of this state is stable as
imprinted also in the time evolution of $M(t)$ for $|B_0|<0.04$ [see Fig.  \ref{fig:SU2viol}(c)]. In
the vicinity of $B_0 \approx 0$, $M(t)$ is depleted during the time-evolution while $C_{FF}(t)
\approx 1$ throughout the dynamics. The appearance of this region is explained by the fact that the
couplings between the states $| \alpha \rangle$, $| \beta \rangle$ and $| \gamma \rangle$ associated
with $\hat{H}_g$ are smaller than their energy differences due to the Anderson kinetic exchange
interaction (being of the order of $\frac{t^b}{g U^b}$). The latter implies a large but finite
life-time of the phase separation of the initial state, in agreement with the SU(2) preserving case
$B=0$. 
In addition, further resonances appear when $|B_0|>0.04$ for $g=2$ [see Fig. \ref{fig:SU2viol}(c)]. More specifically, the
resonances at $B_0>0.04$ correspond to tunneling resonances in a similar fashion to the case of the
interaction regime $A$ [Fig. \ref{fig:SU2viol}(a)]. The positive shift of these resonances when
compared to the corresponding ones appearing for $g=0.5$ is attributed to the increased
interaction energy of the states accessed by tunneling.  For $B_0<-0.04$ another set of resonances
occurs in Fig. \ref{fig:SU2viol}(c) that correspond to interband processes similar to the aforementioned cradle mode.  These
resonances emerge due to the coupling of different bands induced by the interactions.

Within the interaction regime $C$ the stability properties of the phase separation are similar to
the corresponding ones of the interaction regime $A$, compare in particular Fig.
\ref{fig:SU2viol}(e) and \ref{fig:SU2viol}(f) to Fig. \ref{fig:SU2viol}(a) and \ref{fig:SU2viol}(b)
respectively. For large $B_0<0$ the initial state is stable [see Fig.  \ref{fig:SU2viol}(e) and
\ref{fig:SU2viol}(f)], however, for $B_0 \approx 0$ the phase separation and intra-well
ferromagnetic order as imprinted in $M(t)$ and $C_{FF}(t)$ respectively fluctuate during the
dynamics. This fluctuating behaviour can be explained by the inter-band coupling that occurs within
this interaction regime suppressing the intra-well ferromagnetic order of the initially phase
separated state [see also Fig. \ref{fig:comparison}(c) and \ref{fig:comparison}(d)].  Turning to
large $B_0>0$ the phase separation is stable [see in particular Fig.  \ref{fig:SU2viol}(e)] since
$M(t) \approx 1$. However, for $B_0 \approx 0.04$ a resonance associated with the narrow avoided
crossings identified in Fig. \ref{fig:MB_res1}(a) for $g \approx 3$ is observed.

The above discussed stability properties of the phase separated state, especially within the
interaction regime $B$, provide direct insight into the magnetic properties of the SU(2) violating
system. First, the fact that the phase separated state, $| \Psi (0) \rangle$, which is not an
eigenstate of $\hat S^2$, becomes an eigenstate of the system, $\hat H = \hat H_0 +\hat H_I + \hat
H_g$, even for a relatively small breaking of the SU(2) symmetry shows that, as also identified
previously, for a DW there is no global ferromagnetic order imprinted in $S^2$. This is in contrast
to the case of the harmonic confinement as it has been demonstrated in Ref. \cite{Koutent}. Instead,
for a DW trap the instability of the $S^2$ becomes more pronounced for intermediate interactions.
This property can be understood by inspecting the effective tJU model [Eq. (\ref{eq:tJU_model})].
For fermions confined in a DW, ferromagnetic Hund interactions occur {\it only} between particles
that reside in the same well and as a consequence only the intra-well ferromagnetic correlations are
robust within each well. An observation that is also supported by the apparent stability of the
phase separated state except for the cases within the interaction regimes $B$ and $C$ where
inter-band couplings are involved, see Fig. \ref{fig:SU2viol}(b), \ref{fig:SU2viol}(c),
\ref{fig:SU2viol}(d) and \ref{fig:SU2viol}(f). 
Most importantly, for intermediate interactions supporting the intra-well ferromagnetic order [see
Fig. \ref{fig:SU2viol}(c)] the phase separated state is stabilized even for a very weak breaking of the SU(2)
symmetry. This feature of the DW system can be understood by the fact that the extremely weak
Anderson kinetic exchange interaction is the only magnetic mechanism that can possibly prohibit the
coupling of states with different $S$ for a system with broken SU(2) symmetry. On the contrary, the
intra-well ferromagnetic order is stable independently of whether the SU(2) symmetry is preserved or
it is weakly broken as the intra-well ferromagnetic correlations are protected by the much stronger
Hund exchange interaction. The above imply that within the interaction regime $B$ dominated by
ferromagnetic intra-well correlations an instability occurs which is triggered by the
breaking of the SU(2) symmetry. This instability leads to the formation of two polarized
ferromagnetic domains of the spin components as
the system phase separates almost perfectly among the two wells.

\section{Conclusions}
 
We have explored the stability of the phase separated state of interacting spin-$1/2$ fermions
confined in DW potentials. Most importantly, we have revealed an interaction regime characterized
by a metastable phase separation for moderate interactions. 
By invoking an effective tight-binding model, we unveil that the metastability of the phase separation
is related to the formation of a quasi-degenerate manifold of states described by ferromagnetic
intra-well spin-spin correlations but varying total spin. The formation of this quasi-degenerate
manifold of states can be intuitively understood by the inclusion of an effective ferromagnetic Hund
interaction, stemming from the spin exchange interaction between two interacting particles residing
at the same well. This exchange interaction cannot be neglected due to the large spatial overlap of
the particles occupying different bands but the same well of the DW. The breaking of the SU(2) symmetry
is found to substantially alter the behaviour of the system in this interaction regime where the
ferromagnetic correlations dominate. Indeed, the phase separated state becomes stable even when we
break the SU(2) symmetry by employing a very weak linear gradient of the magnetic potential.

The description of the magnetic properties of 1D fermions in terms of the ferromagnetic Hund
interaction provides a unifying viewpoint on the relation between phase separation and
ferromagnetism.  Most importantly, it provides a theoretical framework via which the stability of
ferromagnetic correlations in the absence of SU(2) symmetry
(see also \cite{Koutent}) can be understood. In particular, the ferromagnetic correlations are found
to be stable only within the spatial regions where the Hund interaction is strong, i.e. within each
of the wells of a DW and not between them. In this picture the ferromagnetic correlations of the
system are not directly related with the phase separation in contrast to the conventional Stoner
instability viewpoint. Instead, the effective antiferromagnetism induced by the Anderson kinetic
exchange interaction is responsible for the absence of phase separation in SU(2) symmetric systems.
Indeed, when this effective antiferromagnetism is weak the system is found to be unstable towards
phase separation. 
More precisely, in the case of a DW potential these two phenomena are indeed related. In the
interaction regime where the ferromagnetic Hund interaction dominates the Anderson kinetic exchange
interaction leading to stable intra-well ferromagnetic correlations, even a weak breaking of the
SU(2) symmetry enforces the system to phase separate.

Our work sets several avenues of further study that can be pursued. First, notice the absence of any
obvious limitation of the underlying mechanisms that would make them incapable of describing higher
dimensional settings. The examination of higher dimensional settings is therefore a promising next
step for understanding the ferromagnetic properties emerging in DW systems. Also, the tunability of
the phase separation by weakly breaking the SU(2) symmetry gives rise to the prospect of controlling
the formation of ferromagnetic domains in the case of DW or lattice systems.  Finally, the inclusion
of various inherent effects that break the SU(2) symmetry of a Fermi system such as spin-orbit
coupling or weak spin-dependent interactions might allow cold atoms to form realistic models that
better emulate the ferromagnetic properties encountered in real materials.

\begin{acknowledgements}
    This work is funded by the Cluster of Excellence 'CUI:Advanced Imaging of Matter' of the Deutsche
    Forschungsgemeinschaft (DFG) - EXC 2056 - project ID 390715994. S.I.M.
    gratefully acknowledges financial support in the framework of the Lenz-Ising Award of the
    University of Hamburg. 
\end{acknowledgements}
\appendix
\section{Six Fermion Dynamics} \label{sec:N6}

\begin{figure}[t]
    \includegraphics[width=1.0\textwidth]{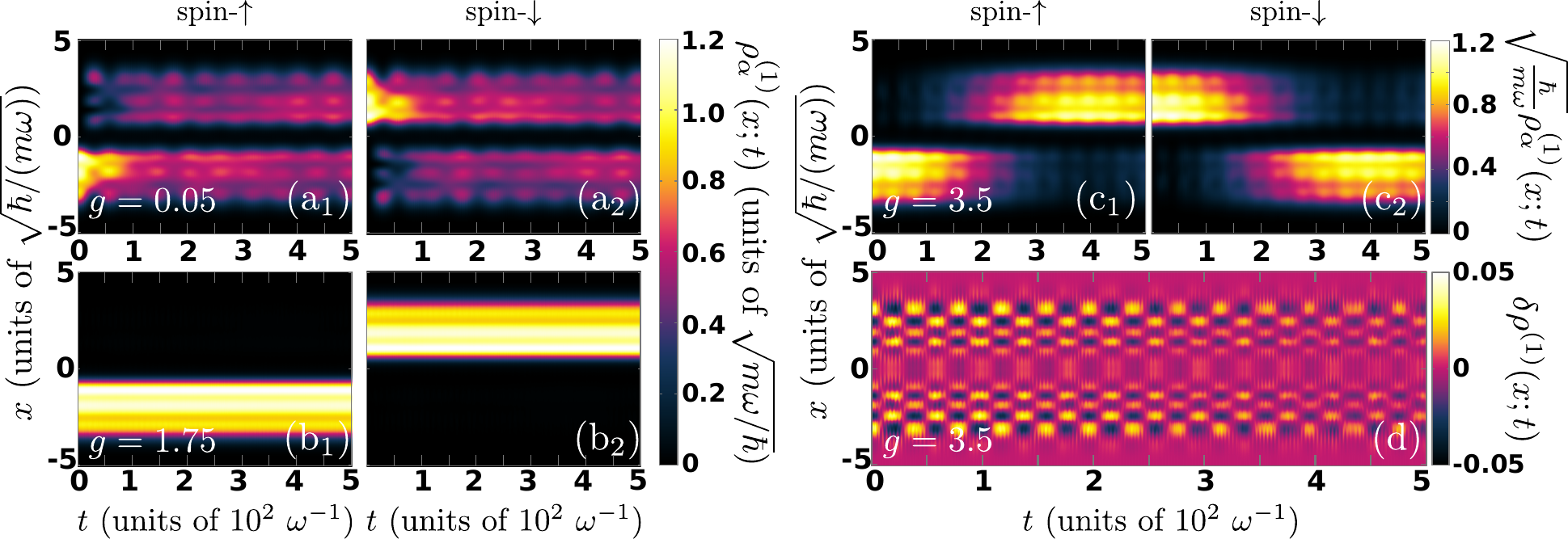}
    \caption{(a$_i$), (b$_i$), (c$_i$) Time-evolution of the one-body densities
        $\rho^{(1)}_{\alpha}(x;t)$ for ($i=1$) the spin-$\uparrow$ and ($i=2$) the spin-$\downarrow$
        component for three different interaction strengths (see legend). (d) The total one-body
        density fluctuations, $\delta\rho^{(1)}(x;t)$ for $g=3.5$, where the cradle mode is clearly
    imprinted. In all cases $w=0.5$, $V_0=10$ and $N_\uparrow=N_\downarrow=3$. }
    \label{fig:MB_res3}
\end{figure}

The discussion in section IV B reveals that within the effective model description an overall
similar dynamical behaviour of the system is expected independently of the particle number [see also
footnote 2]. To verify this expectation within the fully correlated approach we investigate the
dynamical behaviour of a system consisting of $N=6$ fermions and identify the underlying
phenomenology associated with the different interaction regimes $A$, $B$ and $C$ in the
corresponding spin-$\alpha$ one-body densities, $\rho^{(1)}_{\alpha}(x;t)$, illustrated in Fig.
\ref{fig:MB_res3}.

In particular, for weak interactions ($g=0.05$) the one-body density of both spin-$\uparrow$ and
spin-$\downarrow$ fermions exhibits a tunneling dynamics among the wells, see Fig.
\ref{fig:MB_res3}(a$_1$) and Fig. \ref{fig:MB_res3}(a$_2$). In this case, each of the particles
occupying the three energetically lowest bands performs an individual tunneling oscillation with a
frequency close to the one associated with the band it occupies, $t^b$, see for instance the fast
tunneling of the three-humped structure emerging in $\rho^{(1)}_{\alpha}(x;t)$ in comparison to the
overall slower tunneling dynamics. This observed dynamics is in line to the one emerging within the
interaction regime $A$ for $N=4$ particles [compare Fig. \ref{fig:MB_res2}(a) with Fig.
\ref{fig:MB_res3}(a)]. For increasing interactions, $g=1.75$, no tunneling oscillations are observed
and the phase separation appears to be almost completely stable within the time scales we
have studied, see Fig.  \ref{fig:MB_res3}(b$_1$) and \ref{fig:MB_res3}(b$_2$). This behaviour of the
one-body density is characteristic for the interactions belonging to the regime $B$, where as
identified in the $N=4$ particle case the tunneling dynamics slows down dramatically [see also Fig.
\ref{fig:MB_res2}(b)] as a consequence of the formation of the quasi-degenerate manifold of
eigenstates with ferromagnetic intra-well correlations. Finally, the cradle mode being the
characteristic feature of the interaction regime $C$ [see also Fig. \ref{fig:MB_res2}(c) and
\ref{fig:MB_res2}(d)] can also be observed for $N=6$. Inspecting the dynamics of the one-body
density for $g=3.5$ [Fig. \ref{fig:MB_res3}(c)], we observe a collective tunneling mode of the density
among the wells, as well as, deformations of the one-body density within each of the
occupied sites possessing a much larger frequency than the tunneling mode. By employing the temporal
fluctuations of the total one-body density, $\delta \rho^{(1)}(x;t)$ [see Fig.
\ref{fig:MB_res3}(d)] these deformations can be related to the emergence of the cradle mode,
verifying the existence of the interaction regime $C$ in the $N=6$ case.

\section{Shallow Double-Well Case} \label{sec:shallow}

As we have discussed in the main text (see section \ref{sec:corr} and \ref{sec:Su2_viol}) the
relation of the phase separation phenomenon and the ferromagnetism depends on the shape of the
external potential imposed on the atoms. Indeed, it is found that despite the fact that the same
microscopic mechanisms are at play for a parabolically or a DW trapped spin-$1/2$ Fermi system, the
manifestation of the above-mentioned phenomena differs significantly. The purpose of this section is
to study the dependence of the stability properties of the phase separated state, $| \Psi (0)
\rangle$, Eq.  (\ref{initial_state}) on the barrier height of the DW potential. To achieve this we
study the case of a shallower DW with $V_0=5$ and $w=0.5$ and compare with the case of $V_0=8$.
\begin{figure}[t]
    \includegraphics[width=1.0\textwidth]{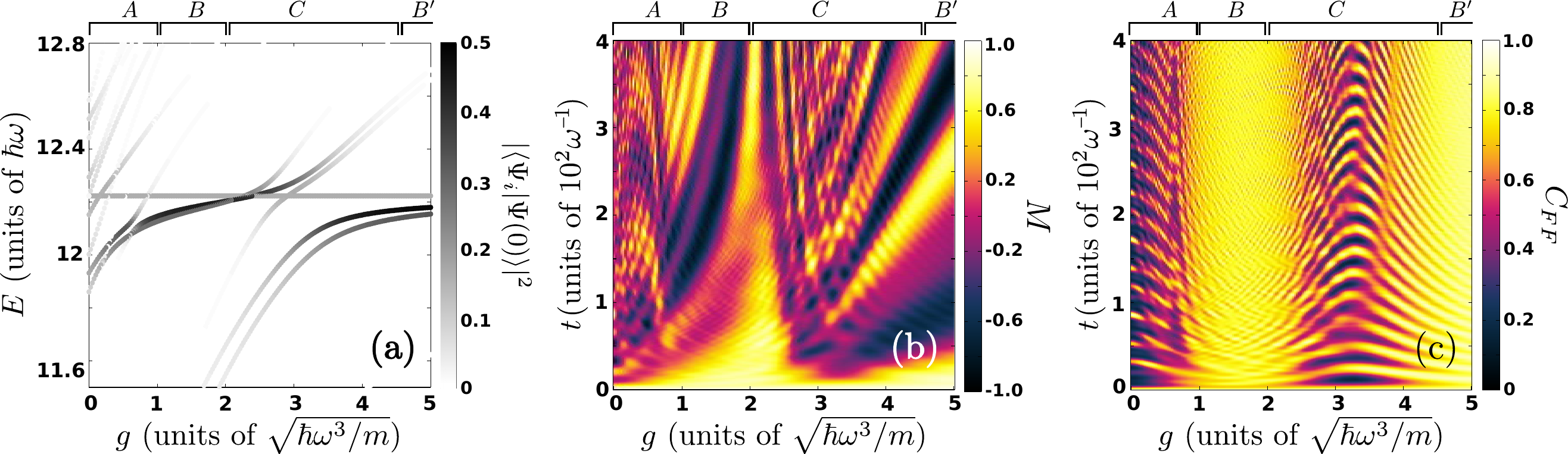}
    \caption{(a) Eigenspectrum of $\hat{H}$ [Eq. (\ref{hamilt})] referring to the eigenstates satisfying the
        overlap criterion $| \langle \Psi_i|\Psi(0)\rangle|^2>10^{-3}$ with
        varying interaction, $g$.  Time-evolution of the (b) magnetization imbalance,
        $M$  and (c) of $C_{FF}$ quanifying the degree of intra-well ferromagnetic correlations for varying $g$. In all cases $w=0.5$, $V_0=5$ and
    $N_\uparrow=N_\downarrow=2$. }
    \label{fig:higherbar}
\end{figure}

The eigenspectrum for a shallow DW is presented in Fig. \ref{fig:higherbar}(a).  The qualitative
structures emerging in the eigenspectrum for $V_0=5$ are similar to the case of $V_0=8$ [compare
Fig. \ref{fig:higherbar}(a) with Fig. \ref{fig:MB_res1}(a)]. However, there are also prominent
quantitative differences which, as we shall explain below, lead to a different dynamical behavior.
Within the regime $A$, $0<g<1$, the role of eigenstates with high energy (see Fig.
\ref{fig:higherbar}(a) for $E>12.2$ and $g<1$) is very pronounced as they accumulate a population
larger than in their deep DW counterpart, see also Fig.  \ref{fig:MB_res1}(a). In the dynamics of
the shallow DW this translates to a much faster loss of $M$ (see e.g. Fig. \ref{fig:higherbar}(b) for
$g<1$) when compared to the case of the deep DW [Fig.  \ref{fig:MB_res1}(b)] which is accompanied
with the loss of intra-well ferromagnetic correlations imprinted in $C_{FF}$, see Fig.
\ref{fig:higherbar}(c). Of course, this difference is simply caused by the larger tunneling rates,
$t^b$ involved in the $V_0=5$ case [Fig. 1(a)]. The differences between the two setups become more interesting
in the intermediate interaction regime, $B$, for $1<g<2.5$. In the shallow DW case three eigenstates
dominate similarly to $V_0=8$, but their spacing is quantitatively larger in the shallower DW
[compare Fig. \ref{fig:higherbar}(a) with Fig.  \ref{fig:MB_res1}(a) for $g \approx 2$]. This is not
surprising since the spacing of these eigenstates (see also section IV B) is proportional to
$\frac{t^b}{g U^b}$ which decreases with increasing $V_0$. In addition, and in direct contrast to
the $V_0=8$ case higher-lying eigenstates [see Fig. \ref{fig:higherbar}(a) for $1<g<2.5$ and
$E>12.3$] and most importantly lower-lying ones [see Fig. \ref{fig:higherbar}(a) for $1<g<2.5$ and
$E>11.8$] are involved in the dynamics within this regime. Accordingly the dynamics of $M$ and
$C_{FF}$ shows that in the shallow DW case the initial state cannot be characterized as metastable
for any interaction in the regime $B$. Indeed, the magnetization imbalance $M(t)$ [Fig.
\ref{fig:higherbar}(b)] is greatly suppressed for $t>100$ possessing values $M(t)<0.6$ for all
interactions in $1<g<2.5$. Regarding the spin-spin correlations it can be seen that $C_{FF}(t)$ is
almost stable during the dynamics except for a very fast decay at initial times $t<4$ [see Fig.
\ref{fig:higherbar}(c)]. During the time-evolution it acquires values of the order of $C_{FF}
\approx 0.8$, for all interactions within the regime $B$ showcasing predominantly ferromagnetic
intra-well correlations. The above implies that while the mechanisms at play in the shallow DW case
are similar to the ones emerging in the case of a deeper DW, the apparent phenomenology is altered
due to the pronounced involvement of lower-lying states.  These lower-lying states are able to alter
the dynamics within the regime $B$ because, as it can be seen by inspecting the eigenspectrum for
$g\approx 3$ the cradle resonances are much wider in the case of a shallower DW thus affecting a
broader interaction regime than for $V_0=8$.

In the case of $V_0=5$ the regime $C$ appears in the
interaction range $2<g<4.5$. The phenomenology taking place within $C$ is completely analogous to
the case of $V_0=8$. Indeed, the tunneling is prevalent within this regime as imprinted in the
fluctuating behavior of the magnetization imbalance $M(t)$ [see Fig. \ref{fig:higherbar}(b)]. In
addition, the intra-well spin-spin correlations imprinted in $C_{FF}(t)$ can be also seen to
fluctuate similarly to the case of $V_0=8$ [compare Fig.  \ref{fig:higherbar}(c) with Fig.
\ref{fig:comparison} (d)]. For even stronger interactions the regime $B'$ is accessed where the
fluctuations of $M$ slow down dramatically when compared to the regions $A$ and $C$ [see Fig.
\ref{fig:higherbar}(b)], while $C_{FF}(t)$ is almost constant during the dynamics possessing values
$C_{FF}(t) \approx 0.9$. In addition, by inspecting the eigenspectrum [Fig.  \ref{fig:higherbar}(a)]
it can be deduced that in this regime a quasi-degenerate manifold of the predominantly occupied
eigenstates begins to form similarly to the regime $B'$ encountered for $V_0=8$ .

In conclusion, the nature of the microscopic mechanisms that govern the stability properties of
phase separation are not altered as the depth of the DW changes. However, because of their direct
competition, in particular between the exchange interaction and the combined effects of the
tunneling and the cradle mode, the observed dynamics differs significantly as the barrier height,
$V_0$ decreases. Indeed, the mechanisms competing with the exchange interaction become more
prevalent for a shallower DW as it is also clearly imprinted in the corresponding eigenspectrum.
This renders the intra-well ferromagnetic order unable to completely dominate the dynamics for every
interaction strength, resulting in the absence of stable ferromagnetic intra-well
correlations and its direct imprint on the dynamics i.e. the metastability of the phase separation.

\section{Anderson Kinetic Exchange Interaction}
The purpose of this section is to provide the explicit derivation of the effective antiferromagnetic
interaction acting upon the different wells of our DW setup.  This antiferromagnetic interaction is
similar to the Anderson kinetic exchange interaction emanating among the different sites of a
lattice within the Hubbard model \cite{Andersonkin}. Although such an effective magnetic term can be
derived within the Resolvent formalism by invoking less assumptions, here we opt to employ the
standard Reyleigh-Schr{\"o}dinger second order perturbation theory due to its mathematical (and
physical) clarity.

The terms appearing in the Hamiltonian of the tJU model [Eq. (\ref{eq:tJU_model})] can be separated
into two Hamiltonian terms\footnote{In order
    to make the notation less cumbersome we drop the index ${\rm eff}$, however all of the
    Hamiltonian terms mentioned in this section are to be considered within the effective tJU
model.} that solely act within each of the wells, $\hat{H}_s$, with $s \in \{ \uparrow, \downarrow
\}$, and a Hamiltonian part corresponding to the
coupling between them, $\hat{H}_{LR}$. By performing this separation the effective Hamiltonian reads
$\hat{H}_{\rm eff}=\hat{H}_R+\hat{H}_L+\hat{H}_{LR}$. The intra-well Hamiltonian terms,
\begin{equation}
    \hat{H}_s=
    +g\sum_{b=0}^{\infty} U^{b} \hat{n}^{b}_{s\uparrow} \hat{n}^{b}_{s\downarrow}
    -g\sum_{b \neq b' \in [0, \infty)} J^{bb'} \left[\hat{\bm{S}}_{s}^{b}\cdot\hat{\bm{S}}_{s}^{b'}
         -\frac{1}{4}  \hat{n}^{b}_{s} \hat{n}^{b'}_{s} \right]\\
    +\sum_{b=0}^{\infty} \epsilon^b \hat{n}^{b}_{s},
\end{equation}
correspond to ferromagnetic Heisenberg models with additional occupation dependent terms $\propto
\hat{n}_s^b$. The
intra-well coupling
\begin{equation}
    \hat{H}_{LR}=-\sum_{b=0}^{\infty}\sum_{\alpha \in \{ \uparrow \downarrow\}} t^b \left( \hat{a}^{b\dagger}_{R\alpha} \hat{a}^b_{L\alpha}
    +\hat{a}^{b\dagger}_{L\alpha} \hat{a}^b_{R\alpha} \right)
\end{equation}
describes the tunneling among the wells. Our intention is to perturbatively treat $\hat H_{LR}$ and
show that it acts as an effective antiferromagnetic interaction between the particles occupying the
same band but different wells.

According to the discussion in sections II B and IV B we are particularly interested in the configuration
with no doublons and a single occupation of each Wannier state up to the $b=\frac{N}{2}-1$ band.
The projection of $\hat{H}_s$ to this particular configuration results in the Heisenberg
model
\begin{equation}
    \hat{P}_B \hat{H}_s\hat{P}_B =\frac{E_B}{2} -g\sum_{b \neq b' \in [0, \frac{N}{2}-1]} J^{bb'} \left[\hat{\bm{S}}_{s}^{b}\cdot\hat{\bm{S}}_{s}^{b'}
    -\frac{1}{4} \right].
\end{equation}
which possesses the degenerate ground states
\begin{equation}
    |\tfrac{N_s}{2},N_{\uparrow s}-\tfrac{N_s}{2}\rangle_s \equiv \sqrt{\frac{(N_s-N_{\uparrow s})!}{N_s!N_{\uparrow s}!}} \left( \sum_{b=0}^{\infty} \hat{S}^b_{+;s} \right)^{N_{\uparrow s}} \left( \prod_{b=0}^{N_s-1} \hat{a}_{s\downarrow}^{b} \right)^{\dagger} | 0 \rangle,
\end{equation}
where $\hat{S}^b_{+;s}\equiv \hat{a}^{b \dagger}_{s \uparrow} \hat{a}^{b}_{s \downarrow}$, $N_L=N_R=\frac{N}{2}$ and we have parametrized these states by the number of spin-$\uparrow$
atoms contained in each well. Then the ground state manifold of the system $\hat{P}_B 
(\hat{H}_L+\hat{H}_R) \hat{P}_B $ possesses an energy $E^{(0)}=E_B=2 \sum_{b=0}^{\frac{N}{2}-1} \epsilon^b$ and it is spanned by the states
$|\Psi;N_{\uparrow L},N_{\uparrow R} \rangle = |\tfrac{N}{4},N_{\uparrow L}-\tfrac{N}{4}\rangle_L
\otimes |\tfrac{N}{4},N_{\uparrow R}-\tfrac{N}{4}\rangle_R$, with $N_{\uparrow L}+N_{\uparrow
R}=\frac{N}{2}$ [see also Eq. (11)].

Note here that the action of $\hat{H}_{LR}$ on the basis of the ground state manifold $|\Psi;
N_{\uparrow L},N_{\uparrow R} \rangle$ is rather simple due to its product state character. Indeed
$\hat{H}_{LR} |\Psi;N_{\uparrow L},N_{\uparrow R} \rangle$ can be expressed via the action of the
creation and annihilation operators on the single-well ferromagnetic states
$|\tfrac{N_s}{2},N_{\uparrow s}-\tfrac{N_s}{2}\rangle_s$.  Indeed the annihilation operator creates
a vacancy to the single-well ferromagnetic states
\begin{equation}
    \begin{split}
    \hat{a}^{b_0}_{s \alpha} \left| \tfrac{N_s}{2}, N_{\uparrow s}-\tfrac{N_s}{2}
    \right\rangle_s=(-1)^{N_s+b_0} \bigg[ \delta_{\alpha \downarrow} \sqrt{\frac{N_s-N_{\uparrow
        s}}{N_s}} \left| \tfrac{N_s-1}{2},  N_{\uparrow s}-\tfrac{N_s-1}{2} \right\rangle_s^{b_0}\\
    + \delta_{\alpha \uparrow} \sqrt{\frac{N_{\uparrow s}}{N_s}} \left| \tfrac{N_s-1}{2},
N_{\uparrow s}-1-\tfrac{N_s-1}{2}\right\rangle_s^{b_0} \bigg].
    \end{split}
    \label{anihfer}
\end{equation}
Nevertheless the resulting states are ferromagnetic since they possess maximal $S_s$ and read
\begin{equation}
    |\tfrac{N_s-1}{2},N_{\uparrow s}-\tfrac{N_s-1}{2}\rangle^{b_0}_s \equiv \sqrt{\frac{(N_s-N_{\uparrow s}-1)!}{(N_s-1)!N_{\uparrow s}!}} \left( \sum_{b=0}^{\infty} \hat{S}^b_{+;s} \right)^{N_{\uparrow s}} \left( \prod_{b=0}^{b_0-1} \hat{a}_{s\downarrow}^{b} \prod_{b=b_0+1}^{N_s-1} \hat{a}_{s\downarrow}^{b} \right)^{\dagger} | 0 \rangle.
    \label{vac}
\end{equation}
Furthermore, the creation operator, $\hat{a}^{b_0 \dagger}_{s \alpha}$, maps the ferromagnetic states to the corresponding
$|\tfrac{N_s-1}{2},N_{\uparrow s}-\tfrac{N_s-1}{2}\rangle^{b_0}_s$ state with an additional doublon
at the $b_0$-th band. More specifically,
\begin{equation}
    \begin{split}
    \hat{a}^{b_0 \dagger}_{s \alpha} \left| \tfrac{N_s}{2}, N_{\uparrow s}-\tfrac{N_s}{2}
    \right\rangle_s=(-1)^{N_s+b_0}\hat{a}^{b_0 \dagger}_{s \uparrow}\hat{a}^{b_0 \dagger}_{s
    \downarrow} \bigg[ \delta_{\alpha \uparrow} \sqrt{\frac{N_s-N_{\uparrow s}}{N_s}} \left|
        \tfrac{N_s-1}{2}, N_{\uparrow s}-\tfrac{N_s-1}{2}  \right\rangle_s^{b_0} \\
    - \delta_{\alpha \downarrow} \sqrt{\frac{N_{\uparrow s}}{N_s}} \left| \tfrac{N_s-1}{2},
N_{\uparrow s} -1 -\tfrac{N_s-1}{2} \right\rangle_s^{b_0} \bigg].
    \end{split}
    \label{creafer}
\end{equation}
Importantly, the intra-well ferromagnetic states with vacancy, $|\tfrac{N_s-1}{2},N_{\uparrow s}-\tfrac{N_s-1}{2}\rangle^{b_0}_s$, are eigenstates of the
Hamiltonian of the corresponding well, $\hat{H}_s$ as they satisfy the following eigenvalue equation
\begin{equation}
    \hat{H}_s |\tfrac{N_s-1}{2},N_{\uparrow s}-\tfrac{N_s-1}{2}\rangle^{b_0}_s=\left(
    \frac{E_B}{2}-\epsilon^b \right) |\tfrac{N_s-1}{2},N_{\uparrow s}-\tfrac{N_s-1}{2}\rangle^{b_0}_s.
\end{equation}
By employing the commutation relations of the creation operator of a doublon $\hat{a}^{b \dagger}_{s
\uparrow}\hat{a}^{b \dagger}_{s \downarrow}$, namely
$[\hat{n}_{s}^b,\hat{a}^{b' \dagger}_{s' \uparrow}\hat{a}^{b' \dagger}_{s' \downarrow}]=2 \delta_{b b'} \delta_{s s'} \hat{a}^{b' \dagger}_{s' \uparrow}\hat{a}^{b' \dagger}_{s' \downarrow}$, 
$[\hat{n}_{s\uparrow}^b \hat{n}_{s\downarrow}^b,\hat{a}^{b' \dagger}_{s' \uparrow}\hat{a}^{b'
\dagger}_{s' \downarrow}]= \delta_{b b'} \delta_{s s'} \hat{a}^{b' \dagger}_{s' \uparrow}\hat{a}^{b'
\dagger}_{s' \downarrow}$ and  
$[\hat{\bm{S}}_{s}^b,\hat{a}^{b' \dagger}_{s' \uparrow}\hat{a}^{b' \dagger}_{s'
\downarrow}]=\bm{0}$, it can be shown that the states containing an additional doublon satisfy the
following eigenvalue equation
\begin{equation}
    \begin{split}
    \hat{H}_s \hat{a}^{b_0 \dagger}_{s \uparrow}\hat{a}^{b_0 \dagger}_{s \downarrow}
    |\tfrac{N_s-1}{2},N_{\uparrow s}-\tfrac{N_s-1}{2}\rangle^{b_0}_s=\left( \frac{E_B}{2} +
    \epsilon^{b_0} + g  \tilde{U}^{b_0} \right) 
    \hat{a}^{b_0 \dagger}_{s \uparrow}\hat{a}^{b_0 \dagger}_{s \downarrow}
    |\tfrac{N_s-1}{2},N_{\uparrow s}-\tfrac{N_s-1}{2}\rangle^{b_0}_s,
    \end{split}
\end{equation}
where $\tilde{U}^{b_0}=\sum_{b=0}^{\frac{N}{2}-1} J^{b_0 b}$ (recall that $U^{b_0}=J^{b_0 b_0}$), and as a consequence also constitute eigenstates of $\hat{H}_s$.

\begin{figure}[t]
    \includegraphics[width=1.0\textwidth]{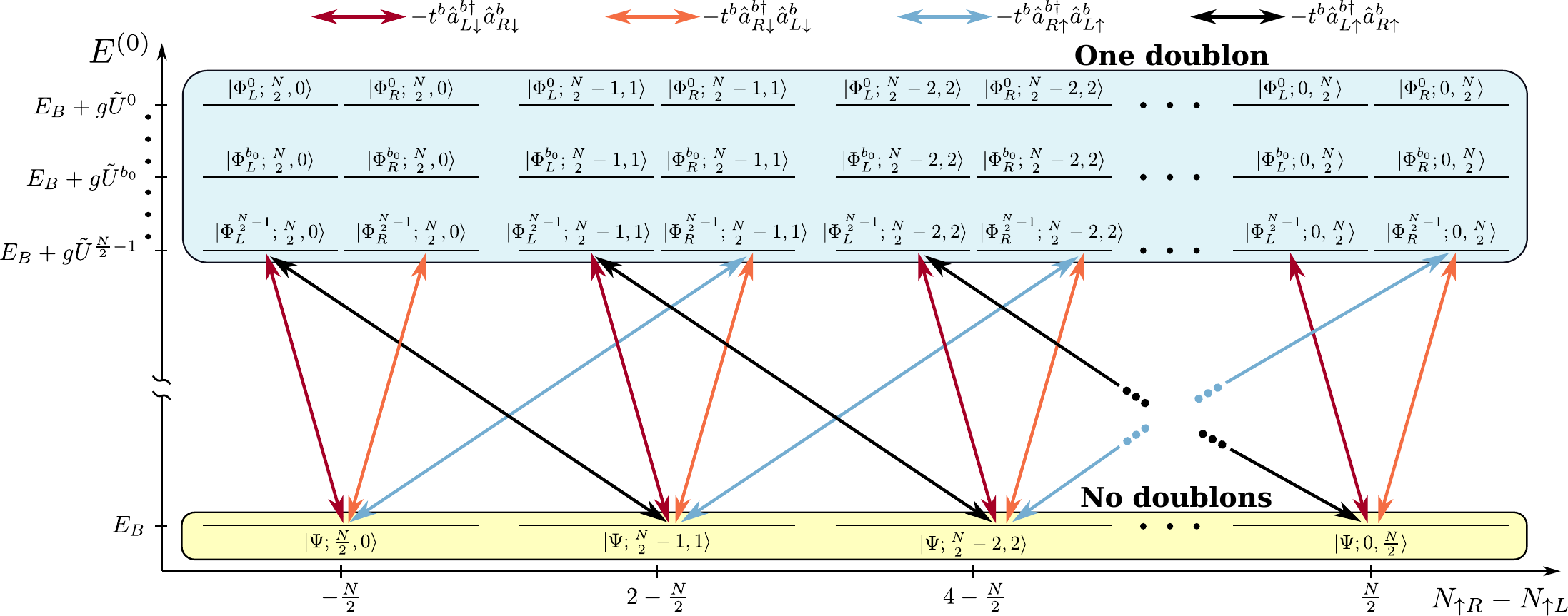}
    \caption{Schematic illustration of the coupling scheme introduced by the action of
        $\hat{H}_{LR}$ on the eigenstates of $\hat{H}_L+\hat{H}_R$ for the configurations with zero
        and one double occupations. The colored arrows indicate the distinct transitions introduced
        by each of the terms appearing in $\hat{H}_{LR}$. For clarity only the transitions
    involving the $b=\frac{N}{2}-1$ band are depicted.}
    \label{fig:coupls}
\end{figure}

By using Eq. (\ref{anihfer}) and (\ref{creafer}) we can show that each term appearing in
$\hat{H}_{RL}$ couples each state $|\Psi;N_{\uparrow L},N_{\uparrow R} \rangle$ to a single state
containing one doublon. This coupling scheme is schematically depicted in Fig. \ref{fig:coupls}.
Here the state $|\Phi^{b_0}_s;N_{\uparrow L}, N_{\uparrow R} \rangle$ refers to the $\vec{n}_B$
state possessing a double occupancy
at the $b_0$-th band of the $s$-well and $N_{\uparrow L}$ and $N_{\uparrow R}$ spin-$\uparrow$ atoms in the left and right
well respectively. For $s=R$ this state possessing a doublon reads $|\Phi^{b_0}_R;N_{\uparrow L}, N_{\uparrow R} \rangle = |\tfrac{N/2}{2},N_{\uparrow
L}-\tfrac{N/2-1}{2}\rangle^{b_0}_L \otimes \hat{a}^{b_0 \dagger}_{R
\uparrow}\hat{a}^{b_0 \dagger}_{R \downarrow} |\tfrac{N/2-1}{2},N_{\uparrow R}-1
-\tfrac{N/2-1}{2}\rangle^{b_0}_R$ and for $s=L$,  $|\Phi^{b_0}_L;N_{\uparrow L}, N_{\uparrow R} \rangle = \hat{a}^{b_0 \dagger}_{L
\uparrow}\hat{a}^{b_0 \dagger}_{L \downarrow}|\tfrac{N/2}{2},N_{\uparrow
L}-1-\tfrac{N/2-1}{2}\rangle^{b_0}_L \otimes  |\tfrac{N/2-1}{2},N_{\uparrow R}
-\tfrac{N/2-1}{2}\rangle^{b_0}_R$. Moreover, it can be shown that $|\Phi^{b_0}_s;N_{\uparrow L}, N_{\uparrow R} \rangle$ are
eigenstates of $\hat{H}_L + \hat{H}_R$ and are degenerate. Indeed,
the following eigenvalue equation holds
\begin{equation}
        \left( \hat{H}_L + \hat{H}_R \right) |\Phi_s^{b_0};N_{\uparrow
        L},N_{\uparrow R} \rangle =\left( E_B + g \tilde{U}^{b_0} \right)
    |\Phi_s^{b_0};N_{\uparrow L},N_{\uparrow R} \rangle.
    \label{Phisb0}
\end{equation}
Therefore, by employing the basis states $|\Psi;N_{\uparrow L}, N_{\uparrow R} \rangle$ and $|\Phi^{b_0}_s;N_{\uparrow L}, N_{\uparrow R} \rangle$ the couplings between states
possessing no and one double occupation induced by $\hat{H}_{RL}$ are intuitive. Indeed, the
tunneling terms $\hat{a}^{b_0 \dagger}_{R \sigma} \hat{a}^{b_0}_{L \sigma}$ ($\hat{a}^{b_0
\dagger}_{L \sigma} \hat{a}^{b_0}_{R \sigma}$) create a double occupancy on the right (left)
well of the $b_0$-th band and shift $N_{\uparrow L}-N_{\uparrow R}$ by two in the case that
$\sigma=\uparrow$.
For instance, the tunneling term $\hat{a}^{b_0 \dagger}_{R \uparrow} \hat{a}^{b_0}_{L \uparrow}$
(blue arrows in Fig. \ref{fig:coupls}) transfers the spin-$\uparrow$ particle of the state $|\Psi;
N_{\uparrow L}, N_{\uparrow
R} \rangle$ from the left to the
right well of the $b_0$-th band resulting in the formation of a double occupancy on the right well
of this band and modifying the occupation of spin-$\uparrow$ particles to $N_{\uparrow L}-1$ for the left and $N_{\uparrow
R}+1$ for the right well.

The approach followed to obtain the dominant perturbative correction to the eigenstates of
$\hat{H}_L+\hat{H}_R$, $| \Psi; N_{\uparrow L}, N_{\uparrow R} \rangle$, in the presence of the coupling term
$\hat{H}_{RL}$ is explicated below. First, we define the Hilbert space spanned by the
degenerate eigenstates $| \Psi; N_{\uparrow L}, N_{\uparrow R} \rangle$ as $\mathcal{H}_0$.
Obviously since the states within $\mathcal{H}_0$ are not directly coupled by $\hat H_{LR}$ the
first order perturbative correction to their energy vanishes. In order to obtain the first
non-trivial correction to the energy of those degenerate states we have to treat the coupling term
$\hat{H}_{LR}$ within second order perturbation theory.  Let us define the perturbative eigenstates
up to second order in perturbation theory as $|S_{i}\rangle \approx |S^{(0)}_{i}\rangle +
|S^{(1)}_{i}\rangle + |S^{(2)}_{i}\rangle$ with $|S^{(0)}_{i}\rangle \in \mathcal{H}_0$.
Accordingly, the corresponding perturbative eigenenergies read, $E_i
\approx E^{(0)} + E_i^{(2)}$, with $E^{(0)}=E_B$. Then, the second order correction to the eigenergy
of $|S_{i} \rangle$, $E^{(2)}_i$ is given by
\begin{equation}
    \langle S^{(0)}_{i}| \sum_{k \notin \mathcal{H}_0} \frac{\hat{H}_{LR} | \Psi_k \rangle \langle
    \Psi_k | \hat{H}_{LR}}{E^{(0)}-E^{(0)}_k} |S^{(0)}_{j}\rangle =\delta_{ij} E^{(2)}_{i},
   \label{correction}
\end{equation}
where $| \Psi_k \rangle$ and $E^{(0)}_k$ correspond to the eigenstates and eigenenergies of
$\hat{H}_L+\hat{H}_R$ within the complementary space of $\mathcal{H}_0$.  Eq. (\ref{correction})
implies that in order to obtain $E^{(2)}_i$ the operator in the
braket should be diagonalized within $\mathcal{H}_0$. The eigenvectors resulting from this
diagonalization correspond to the zeroth order correction of the eigenstate, $|S^{(0)}_i \rangle$.
Having found the zeroth order correction to the eigenstate the first order correction to it can be
derived by employing
\begin{equation}
    |S^{(1)}_{i}\rangle =\sum_{k \notin \mathcal{H}_0} \frac{\langle \Psi_k | \hat{H}_{LR} | S^{(0)}_i \rangle}{E^{(0)}-E^{(0)}_k} |\Psi_k\rangle
\end{equation}
and finally the second order correction can be similarly obtained by
\begin{equation}
    |S^{(2)}_{i}\rangle =\sum_{m \notin \mathcal{H}_0} \langle \Psi_m | \sum_{k \notin \mathcal{H}_0} \frac{\hat{H}_{LR} | \Psi_k \rangle \langle \Psi_k | \hat{H}_{LR}}{(E^{(0)}-E^{(0)}_m)(E^{(0)}-E^{(0)}_k)} | S^{(0)}_i \rangle |\Psi_m\rangle.
\end{equation} 

The fact that the states $|\Phi_s^b; N_{\uparrow L}, N_{\uparrow R} \rangle$ which are coupled to
the $\mathcal{H}_0$ manifold [see Fig. \ref{fig:coupls} and Eq. (\ref{Phisb0})] are degenerate greatly simplifies Eq.
(\ref{correction}). Indeed the latter can be expressed as
\begin{equation}
    \langle S^{(0)}_{i}| \sum_{b=0}^{\infty} \sum_{\alpha,\alpha' \in \{\uparrow \downarrow\} }
    \frac{2 (t^b)^2}{g \tilde{U}^b}  \left( \hat{a}_{L \alpha}^{b \dagger} \hat{a}_{R \alpha}^b \hat{a}_{R \alpha'}^{b \dagger} \hat{a}_{L \alpha'}^b + \hat{a}_{R \alpha}^{b \dagger} \hat{a}_{L \alpha}^b \hat{a}_{L \alpha'}^{b \dagger} \hat{a}_{R \alpha'}^b \right) |S^{(0)}_{j}\rangle =\delta_{ij} E^{(2)}_{i}
    \label{<+label+>}
\end{equation}
and by introducing the spin-operators $\hat{\bm S}^b_s$ an effective antiferromagnetic Heisenberg exchange interaction term is obtained
\begin{equation}
    \langle S^{(0)}_{i}| \sum_{b=0}^{\infty} \frac{4 (t^b)^2}{g \tilde{U}^b}  \left( \hat{\bm{S}}^b_L \cdot \hat{\bm{S}}^b_R - \frac{1}{4} \right) |S^{(0)}_{j}\rangle =\delta_{ij} E^{(2)}_{i}.
\end{equation}
Obviously, this Heisenberg exchange interaction term possesses the $S^2$ [SU(2)] symmetry and as a
consequence the zeroth order correction $|S^{(0)}\rangle$ can be identified with the states of
definite $S$, $|\Phi(t^b=0);S \rangle \in \mathcal{H}_0$. Then the first order $|S^{(1)}\rangle$
and the second order $|S^{(2)}\rangle$ corrections to the wavefunction correspond to the occupation
of states possessing one and two double occupations respectively. In the limit $\frac{t^b}{g U^b}
\ll 1$ the occupation of these states becomes highly suppressed and as a consequence these
corrections can be neglected.  Indeed, as Fig. \ref{fig:comparison} (b) reveals such corrections
even beyond the effective tJU model contribute to a correction less than $2\%$ to the fully
correlated many-body eigenstates within the interaction regime $B$. Within the above mentioned
approximation the $\hat H_{RL}$ coupling term can then be substituted with the one of the effective Anderson
exhange interaction
\begin{equation}
    \hat{H}_{RL} \approx \hat{H}_{RL}^{\rm And}=\sum_{b=0}^{\infty} \frac{4 (t^b)^2}{g \tilde{U}^b}  \left( \hat{\bm{S}}^b_L
    \cdot \hat{\bm{S}}^b_R - \frac{1}{4} \right),
\end{equation}
which corresponds exactly to the form of the effective antiferromagnetic interaction appearing in 
Eq. (10) of the main text.

\section{The Computational Method: ML-MCTDHF}

To solve the MB Schr\"{o}dinger equation $\left( {i\hbar {\partial _t} - \hat H} \right) |\Psi (t)
\rangle= 0$ we rely on the multilayer multiconfiguration time-dependent Hartree method for atomic
mixtures \cite{MLX} (ML-MCTDHX). More specifically, a reduction of the ML-MCTDHX method for
spin-$1/2$ fermions is employed which is referred to as the spinor-variant of the multiconfiguration
time-dependent Hartree method for Fermions (MCTDHF).  MCTDHF has been applied extensively for the
treatment of fermions with or without spin-degrees of freedom, in a large class of condensed matter,
atomic and molecular physics scenarios (see e.g.
\cite{Zanghellinisup,Katosup,Caillatsup,Nestsup,Bonitzsup,Haxtonsup}) and recently also applied in
the field of ultracold atoms \cite{Koutent,MLX,Jenny,Jenny1,Jens,Fpolarons,CaoMistakidis}. MCTDHF
is a variational method the key idea of which is to employ a time-dependent (TD) and variationally
optimized MB basis set, which allows for the optimal truncation of the MB Hilbert space.  The
ansatz of the MCTDHF method can be summarized as follows. First, the MB wavefunction, $| \Psi(t)
\rangle$ is expanded on a TD number-state basis
\begin{equation}
    | \Psi (t) \rangle =\sum_{\vec{n}} A_{\vec{n}}(t) | \vec{n} (t) \rangle,
\label{eq:mb_wfn}
\end{equation}
where $A_{\vec{n}}(t)$ are the corresponding TD expansion coefficients.
The TD number states $|\vec{n} (t) \rangle$ each possessing different occupation numbers
$\vec{n}=(n_1,\dots,n_D)$ read
\begin{equation}
    | \vec{n}(t) \rangle = \left[ \prod_{i=1}^D \hat{a}^{n_i}(t) \right]^{\dagger} |0 \rangle.
    \label{eq:ns}
\end{equation}
As Eq. (\ref{eq:ns}) reveals the time-dependence of this MB basis stems from the utilization of $D$
different TD creation operators,
$\hat{a}^{\dagger}_j(t)$, $j=1,\dots,D$. These operators create a fermion in 
the TD and variationally optimized single particle function (SPF)
\begin{equation}
    | \phi_j (t) \rangle= \hat{a}^{\dagger}_j(t) | 0 \rangle = \int {\rm d}x~\left[ \phi_{j\uparrow}(x;t)
    \hat{\psi}^{\dagger}_{\uparrow}(x)+\phi_{j\downarrow}(x;t)
    \hat{\psi}^{\dagger}_{\downarrow}(x)\right] | 0 \rangle,
\label{eq:spfs}
\end{equation}
where the variational parameters $\phi_{j \alpha}(x;t)$ refer to the spatial distribution of
the spin-$\alpha$ part of the $j$-th SPF and $\hat{\psi}_{\alpha}(x)$ is the spin-$\alpha$ fermionic
field operator. The operators $\hat{a}_j(t)$ satisfy the standard fermionic anti-commutation
relations $\{ \hat{a}_i(t) , \hat{a}^{\dagger}_j(t) \}=\delta_{i,j}$ and thus the MCTDHF ansatz
takes explicitly into account the particle symmetry of the system.  Note here that we have used the
term spinor-variant when referring to our implementation of MCTDHF as each SPF, $| \phi_j (t)
\rangle$, in our case is a general spinor wavefunction [see Eq.  (\ref{eq:spfs})].  By employing the
above mentioned ansatz Eq. (\ref{eq:mb_wfn}), (\ref{eq:ns}) and (\ref{eq:spfs}) the time-evolution
of the $N$-body wavefunction, $| \Psi(t) \rangle$ under the effect of the Hamiltonian $\hat{H}$
reduces to the determination of the coefficients $A_{\vec{n}}(t)$ and the components of the SPFs,
$\phi_{j\uparrow}(x;t)$, $\phi_{j\downarrow}(x;t)$. The latter in turn follow the variationally
obtained MCTDHF equations of motion \cite{MLX}. In the limiting case of $D=N$, the method reduces to
the time-dependent Hartree-Fock approach neglecting all two-body and higher-order correlations. In
the opposite limiting case of $D=2 M_{p}$, where $M_{p}$ is the dimension of the basis for the SPF
coefficients, MCTDHF is equivalent to a full configuration interaction approach (commonly referred
    to as ``exact diagonalization'' in the literature). The major advantage of the MCTDHF method
    when compared to methods employing a stationary single-particle basis is that the employed
    time-dependent basis is able to adapt to the correlation patterns emerging in the system during
    the dynamics and thus a smaller set of basis states is required for numerical convergence.

For our implementation we discretize the spatial coordinate by employing a harmonic oscillator
discrete variable representation (DVR), which results after a unitary transformation of the commonly
employed basis of harmonic oscillator eigenfunctions. To study the dynamics, we propagate the
wavefunction by utilizing the appropriate Hamiltonian within the MCTDHF equations of motion.  To
verify the accuracy of the numerical integration, we impose the following overlap criteria $\left|
\langle \Psi | \Psi \rangle - 1 \right| < 10^{-8}$ for the total wavefunction and $\left| \langle
\phi_i | \phi_j \rangle - \delta_{ij} \right| < 10^{-9}$  for the SPFs.  To testify convergence, we
increase the number of SPFs and DVR basis states such that the observables of interest ($M$,
$C_{FF}$) do not change within a given level of accuracy which is in our case $10^{-4}$. More
specifically, we have used $M_{p}=80$, $D=16$ and $M_{p}=80$, $D=18$ for the $N=4$ and the $N=6$
case respectively.  Note that a full configuration interaction treatment of the above-mentioned
systems in the employed primitive bases would require $2.63\times 10^7$ number states for $N=4$ and
$2.12\times 10^{10}$ ones for $N=6$.

{}

\end{document}